\documentclass[11pt]{article}

\usepackage{jheppub}

\usepackage{amsmath, amssymb}
\usepackage[english]{babel}
\usepackage[utf8]{inputenc}
\usepackage{enumitem}
\usepackage{latexsym}
\usepackage{graphicx}
\usepackage{slashed}
\usepackage{xcolor}
\usepackage{color}
\usepackage{float}
\usepackage{bm}
\usepackage{braket}
\usepackage{dsfont}
\usepackage{cancel}
\usepackage{mathrsfs}
\usepackage{tikz}

\def\simgt{\mathrel{\lower2.5pt\vbox{\lineskip=0pt\baselineskip=0pt
           \hbox{$>$}\hbox{$\sim$}}}}

\newcommand{\postscript}[2]{\setlength{\epsfxsize}{#2\hsize}
  \centerline{\epsfbox{#1}}}

\usepackage{caption}
\usepackage{subcaption}

\usepackage{hhline}

\usepackage{array}

\newcolumntype{C}[1]{>{\centering\arraybackslash}p{#1}}

\title{\bf Cosmological history after higher dimensional inflation}

\author[a,b,c]{Luis A. Anchordoqui,}

\affiliation[a]{Department of Physics and Astronomy,  Lehman College, City University of
  New York, NY 10468, USA
}

\affiliation[b]{Department of Physics,
 Graduate Center,  City University of
  New York,  NY 10016, USA
}

\affiliation[c]{Department of Astrophysics,
 American Museum of Natural History, NY
 10024, USA
}

\author[d,e]{Ignatios Antoniadis,}

\affiliation[d]{School of Natural Sciences, Institute for Advanced Study, Princeton, NJ 08540, USA}

\affiliation[e]{Laboratoire de Physique Th\'eorique et Hautes \'Energies
  - LPTHE 
Sorbonne Universit\'e, CNRS, 4 Place Jussieu, 75005 Paris, France
}

\author[f]{and Jules Cunat,}

\affiliation[f]{High Energy Physics Research Unit, Faculty of Science, Chulalongkorn University, Bangkok 1030, Thailand}

\abstract{It was proposed that extra dimensions can acquire large size by higher dimensional inflation connecting two large hierarchies in particle physics and cosmology, namely the weakness of the actual gravitational force to the largeness of the observable universe, in terms of one fundamental scale. This proposal is consistent with the observed approximate scale invariant power spectrum of primordial density perturbations only for one or two extra dimensions of around the micron size. Assuming a stabilisation mechanism of the extra dimensions at the end of inflation, here we propose a cosmological history that describes the Universe evolution after the end of inflation up to the reheating temperature, that guarantees the absence of bulk gravitons at earlier times, avoiding their overproduction in the early universe. The proposed cosmological history connects the period of higher dimensional inflation to the beginning of the standard cosmology.
}

\begin{document}
\maketitle

\section{Introduction}

It is well-known that models featuring large-extra dimensions can
explain significant scale hierarchies in particle physics and
cosmology~\cite{Antoniadis:1990ew,Arkani-Hamed:1998jmv,Antoniadis:1998ig,Montero:2022prj}. Within
these models the Standard Model (SM) fields are localized on a
non-compact four-dimensional (4D) subspace (a.k.a. the brane-world),
while gravity spills 
into a compactified higher-dimensional bulk. For example, the dark
dimension scenario provides a framework of addressing the cosmological
hierarchy problem $\Lambda/M_p^4 \sim 10^{-120}$ by
linking the compact space radius $R_\perp$ to the dark energy scale
$\Lambda$ through the anti-de Sitter (AdS) distance conjecture in de
Sitter (dS) space~\cite{Lust:2019zwm},
\begin{equation}
  R_\perp \sim \lambda_{\rm dd} \Lambda^{-1/4} \,,
\label{rperp}
\end{equation}  
where $M_p$ is the reduced Planck mass and the proportionality factor
is estimated to be $10^{-4} < \lambda_{\rm dd} <
10^{-1}$~\cite{Montero:2022prj}. It is easily seen that if the compact
space has a micrometer-scale characteristic length, then the
Kaluza-Klein (KK) graviton tower necessarily opens up at the mass scale $m_{\rm KK} \sim 1/R_\perp \sim {\rm eV}$ and the higher dimensional $(4+d)$D Planck scale $M_*$ (or species scale where gravity becomes strong~\cite{Dvali:2007hz,Dvali:2007wp,vandeHeisteeg:2022btw,Cribiori:2022nke}) 
is given by\footnote{We consider extra dimensions compactified on line intervals of size $\pi R_\perp$.}
\begin{equation}
  M_* \sim \left(m_{\rm KK}/\pi\right)^{d/(2+d)} \  M_p^{2/(2+d)} \, ,
\end{equation}  
where $d$ is the number of large-extra dimensions. 
It follows that $M_*$ is of order $\sim 10^9$ GeV for $d=1$ and $\sim
10$ TeV for $d=2$, which are the only two possibilities consistent
with collider
experiments and astrophysics
observations~\cite{Anchordoqui:2025nmb,Hannestad:2003yd,Hardy:2025ajb}. 

The dark dimension scenario is constrained from tests of Newtonian
gravity's inverse square law. Indeed, Newton's law has been precisely
verified down to distances of roughly $30~\mu {\rm m}$ in the case of
a circle, or up to $40~\mu {\rm m}$ in case of an
interval~\cite{Lee:2020zjt,Tan:2020vpf}.\footnote{Short-range gravity tests were first proposed to distinguish between these two geometries in~\cite{Schwarz:2024tet}, with current bounds derived in~\cite {Anchordoqui:2026hys}.} Moreover, astrophysics bounds on supernova cooling are  satisfied trivially for $d=1$, but also for $d=2$
when $R_\perp$ is less than a micron~\cite{Hannestad:2003yd,Hardy:2025ajb}.
Conversely,
cosmological data sets restrict the parameter space of the model. In
particular, the thermal overproduction of KK graviton modes in the
early universe sets an upper limit on the ``normalcy'' temperature at which the
Universe must be free of bulk modes~\cite{Arkani-Hamed:1998sfv}. For $d=1$, the normalcy
temperature is estimated to be $T_N \lesssim 100 \, M_p (\Lambda/M_p^4)^{1/6}
\sim 1~{\rm GeV}$~\cite{Gonzalo:2022jac,Anchordoqui:2025opy}. For
$d =2$, the $T_N
\lesssim 4~{\rm MeV}$ constraint~\cite{Hall:1999mk,Hannestad:2001nq} suggests borderline experimental feasibility~\cite{Anchordoqui:2025nmb}.

Besides, for dS spacetime (of radius $1/H$),  there exists
an absolute minimum for the mass of a spin-$2$ particle $m_2$ set by the
Higuchi bound, $m_2^2 \geq  2  H^2$, where $H^2 = \Lambda/(3M_p^2)$ is the Hubble parameter~\cite{Higuchi:1986py}. If
  the bound is violated the massive spin-2 field contains helicity
  modes with negative norm, which are in conflict with
  unitarity. Hence, if we define the ``size'' of the extra dimensions
  by the inverse mass of the lightest KK excitation of the graviton,
  the Higuchi bound forbids any compactification in which the extra
  dimensions are larger than an ${\cal O}(1)$ factor times
  $1/H$~\cite{Kleban:2015daa}.

  Now, since the present value of $H_0 \sim 10^{-34}~{\rm eV} \ll m_{\rm KK} \sim
  \Lambda^{1/4}/\lambda_{\rm dd}$, the Higuchi
bound is inordinately satisfied today by a brane-world accelerated expansion at 
fixed compact space of micron size. However, 
the Higuchi bound forbids the presence of a graviton tower over a dS
background within the mass range $0 \leq m_{\rm KK}^2 \leq 2 H_I^2$, where
$H_I$ is the Hubble parameter during inflation.
The Higuchi bound then forces a very small $H_I \lesssim {\rm eV}$ for brane inflation at fixed
large (micron scale) extra dimension~\cite{Anchordoqui:2022svl}.\footnote{This bound was already imposed in~\cite{Dvali:1998pa} using a different argument.} This type of brane inflationary scenarios also requires fine tuning to accommodate a ridiculously small slow-roll parameter $\varepsilon \sim
10^{-33}$. Naturally, these obstacles could be overcome adopting the
working hypothesis that the Universe undergoes a period of inflation in
which the radius of the dark dimension expanded exponentially fast,
from the species length $R_0 \sim M_*^{-1}$ up to the
micron-scale~\cite{Anchordoqui:2022svl,Anchordoqui:2023etp}.\footnote{We note in passing that inflation with a growing fifth
dimension within a Randall-Sundrum set up~\cite{Randall:1999ee} has been recently explored in~\cite{Mishra:2025ofh}.}
 
Along this line, it has been pointed out that $N$ e-folds of
expansion in the non-compact space emerge from $2N/(2+d)$ e-folds of
the internal radius expansion. It is straightforward to check that expansion from $R_0$ to
$R_\perp \sim 1~\mu{\rm m}$ requires 42 e-folds for $d=1$, and this
corresponds to 63 e-folds in the non-compact space. A similar argument works also for $d=2$~\cite{Anchordoqui:2023etp}. 
Thus, higher dimensional inflation relates two large hierarchies in particle physics and cosmology: 
the weakness of 4D gravity compared to the other interactions  and the largeness of the observable universe which should be causally connected to explain its homogeneity, in terms of a single fundamental scale of an underlying theory of gravity.

Besides, a
period of higher-dimensional slow-roll inflation predicts an
approximate scale invariant power spectrum of primordial density
perturbations, matching cosmic microwave background (CMB)
observations~\cite{Anchordoqui:2023etp}. This is because the two-point
function of a massless, minimally coupled scalar field in dS space
scales logarithmically beyond the cosmological horizon, a property
valid across any spacetime
dimensionality~\cite{Ratra:1984yq}. However, in models with
compactified dimensions, this behaviour typically holds only at scales
smaller than the compactification length; at larger distances, it
deviates from scale invariance, which may conflict with observed
large-scale CMB data. It has been shown
elsewhere~\cite{Anchordoqui:2023etp,Anchordoqui:2024amx} that for
$d=1$, the model's predictions are consistent with {\it Planck}
data~\cite{Planck:2018vyg} and stay within the cosmic variance error band. Below, in Secs.~\ref{setup} - \ref{history}, we restrict our analysis to $d=1$ while the  
$d=2$ case is discussed later in Sec.~\ref{d=2}.

More specifically, at large angular scales  (low-$\ell$ multipole moment 
range) the number of available independent modes to sample $(2 \ell
+1)$ is small, making the fundamental statistical uncertainty inherent
to observing only one universe (cosmic variance) the limiting
factor. 
The angular scale $\theta$ (in degrees) is roughly related to the multipole $\ell$ by the approximation $\theta \approx
  180^\circ/\ell$. Using this approximation, $\ell =10$ to $\ell = 20$
  is a common rule of thumb for 10-degree features.  
  For E-mode polarisation, the spectrum is not typically cosmic
variance limited until very low-$\ell$, but for temperature
anisotropy, the $\ell < 30$ range is considered ``cosmic variance
limited.'' On the other hand, CMB temperature power spectrum measurements suffer from
increasing uncertainty at $\ell \lesssim 10$. In the case of
higher-dimensional inflation, however, a recent analysis of {\it Planck's} TT,
TE, and EE CMB power spectra~\cite{Planck:2018nkj} (accounting for the cosmic variance at
low $\ell$) showed that, at 95\%~CL, the nearly scale-invariant spectrum must be maintained down to $\ell \simeq 3$~\cite{Petretti:2024mjy}. This requirement arises because the so-called low-$\ell$ anomalies are primarily characterised by a suppression of power at $\ell \lesssim 10$, whereas higher-dimensional inflation generically predicts an enhancement of power on these large angular scales. 

Now, $\ell\simeq3$ corresponds to about 27 Gpc
today. Tracing back from the current 3~K universe to a reheating
temperature of $T_r \sim 1~{\rm GeV} $ in a matter- and then radiation-dominated eras is equivalent to looking about 30 e-folds
into the past~\cite{Kreling,German:2020kdp}, at which distances of $27~{\rm Gpc}$ corresponded to roughly $10^{14}~{\rm m}$. However, at the end of inflation the size of the compact space was
around a micron. Therefore, to
ensure that the 5D inflationary  model aligns
with observations, it must include a phase that expands the Universe
from $10^{-6}$ to $10^{14}~{\rm m}$, while suppressing production of bulk
modes. In this paper, we aim to achieve a characterisation of this phase.

The layout is as follows. In Sec.~\ref{sec:normalcytemp} we introduce the normalcy temperature, which consists here in an uper bound on the reheating temperature. In Sec.~\ref{setup}, we summarise the framework of 5D inflation and its 4D description.
In Sec.~\ref{sec:3}, we study the experimental constraints on the
scale of 5D inflation: {\it (i)}~we examine the power
spectrum of cosmological fluctuations to derive upper bounds on the
inflationary Hubble horizon $H_I$ and energy scale of 5D 
inflation $M_I$; {\it (ii)} we impose scale invariance at
wavelengths less than 27 Gpc today by introducing two phases in the
Universe evolution after inflation and before the reheating
temperature; one dominated by inflaton oscillations around the minimum
of its potential when the expansion rate of the Universe was less than
the inflaton mass and another above, up to the end of 5D inflation,
that we parametrise by its number of e-folds and that we will analyse
later; {\it (iii)}~we derive a constraint imposed by solving the
horizon problem to ensure the homogeneity of the
Universe. Section~\ref{sec:4} is devoted to a phenomenological
description of low-temperature reheating, with dominant production of
matter on the brane while suppressing decay of the inflaton into bulk
gravitons. In Sec.~\ref{history}, we put all constraints together and
study the phase between the end of inflation and the start of
oscillations. Finally, in Sec.~\ref{d=2}, we generalize the above
analysis for the case of $d=2$ micron-size extra dimensions. The paper
wraps up in Sec.~\ref{sec:6} with some conclusions and contain three
Appendices that treat the impact of KK decays on estimates of the
relic density, the equivalence between Einstein and Jordan frames for the brane observers, as well as the transiton scale derived from~\cite{Petretti:2024mjy}.

\section{The normalcy temperature\label{sec:normalcytemp}}

A generic feature of models with large extra dimensions is the presence of a large number of KK graviton states coupled to matter on the brane. Although the coupling of each individual mode is suppressed by the four-dimensional Planck scale, the enormous multiplicity of accessible KK states can lead to substantial graviton production in the early universe. If these modes are stable or sufficiently long-lived, they survive as relics and may overclose the Universe. This consideration motivates the introduction of a normalcy temperature $T_N$, defined as the highest temperature at which the bulk remains essentially unpopulated and conventional four-dimensional cosmology is preserved. Since KK gravitons are thermally produced by scattering processes in the primordial plasma, the requirement that their relic abundance remain cosmologically acceptable translates into an upper bound on the reheating temperature after inflation, $T_r\leq T_N$. The total thermal production width of KK gravtiton from the collisions in the thermal plasma is given by
\begin{align}\label{eq:gammabath}
    \Gamma_{\text{KK}}\sim\frac{T^3}{M_p^2}\delta n
\end{align}
where the multiplicity of the KK modes is given by $\delta n=T R_\perp$. Although KK gravitons are produced with typical energies $E_{\text{KK}}\sim T$, the contribution to the KK energy density is dominated by modes with masses $m_{\text{KK}}=\mathcal{O}(T)$. Such modes are therefore only mildly relativistic at production, with $E_{\text{KK}}/m_{\text{KK}}=\mathcal{O}(1)$. Since their physical momentum redshifts as $a^{-1}$ while their mass remains constant, they rapidly become non-relativistic as the Universe expands. For the purpose of estimating their relic abundance, the KK graviton population can therefore be treated as a pressureless matter component, so that
\begin{align}
    \dot{\rho}_{\text{KK}}+3H\rho_{\text{KK}}=\frac{1}{a^3}\frac{\partial}{\partial t}\big(a^3\rho_{\text{KK}}\big)=\Gamma_{\text{KK}}\rho_\text{bath}.
\end{align}
Since KK gravitons are thermally produced by the primordial plasma, we assume that their abundance is generated predominantly during the radiation-dominated (RD) epoch, so that we can further write\footnote{We will neglect the variations of $g_*$ and $g_{*s}$ here for simplicity.}
\begin{align}
    dt=-\sqrt{\frac{90}{\pi^2 g_*}}\frac{M_p}{T^3}dT\qquad\text{and}\qquad\rho_{\text{bath}}=\frac{\pi^2 g_*}{30}T^4,
\end{align}
where $g_*$ ($g_{*s}$) denotes the effective numbers of relativistic degrees of freedom contributing to the energy density (to the entropy density) of the thermal plasma, and using entropy conservation to express the scale factor as
\begin{align}
    a(T)=a_t\left(\frac{g_{*s,t}}{g_{*s}}\right)^{1/3}\frac{T_t}{T}
\end{align}
we can finally estimate the energy density of the KK gravitons today to be
\begin{align}
    \rho_{\text{KK},t}=\frac{1}{a_t^3}\int_{t_r}^{t_t}dt\ a^3\Gamma_{\text{KK}}\rho_{\text{bath}}\simeq\sqrt{\frac{\pi^2 g_{*,r}}{90}}\frac{g_{*s,t}}{g_{*s,r}}\frac{R_\perp T_t^3 T_r^3}{M_p},
\end{align}
where the subscripts $t$ and $r$ respectively denote \emph{evaluated
  today} and \emph{evaluated at reheating}.\footnote{While the
  subscripts $0$ is often preferred than $t$, we will use to express
  initial quantities at the beginning of inflation here.} Assuming
that the KK gravitons are stable on cosmological timescales (validation in Appendix~\ref{App:T_N}), the requirement that their relic population does not overclose the Universe implies
\begin{align}
    \Omega_{\text{KK},t}=\frac{\rho_{\text{KK},t}}{\rho_{c,t}}\leq1
\end{align}
where $\rho_{c,t}$ is the critical energy density today $\rho_{c,t}=3M_p^2H_t^2$. This therefore define the normalcy temperature $T_N$ which is the upper bound on the reheating temperature $T_r$ as
\begin{align}\label{eq:normalcytemp}
    \frac{T_r}{\text{GeV}}\leq \frac{T_N}{\text{GeV}}\sim\left(\frac{R_\perp}{\mu\text{m}}\right)^{-1/3}.
\end{align}
Especially, for a micron-sized extra dimension,\footnote{We note that
  for larger extra dimension, i.e. $R_\perp\sim40\ \mu$m, the normalcy
  temperature found here is very close to the QCD phase transition
  $T_{\text{QCD}}\sim150$ MeV, so that corrections coming from the
  variations of $g_*$ and $g_{*s}$ might become important.} the
reheating temperature has to be lower than $T_N \sim 1~{\rm GeV}$.

\section{Setup of 5D inflation and its four-dimensional description}
\label{setup}

5D inflation with a compact dimension is described by an approximate dS$_5$ metric in which both the compact and the non-compact space dimensions expand exponentially with the 5D proper time. Its line element is:
\begin{align}
\label{dS5}
    ds^2=-d\hat{t}^2+\hat{a}(\hat{t})^2d\vec{x}^2+R(\hat{t})^2dy^2 \quad;\quad R(\hat{t})=R_0\,\hat{a}(\hat{t}) =R_0\,e^{H_I\hat{t}}\,,
\end{align}
where hats are here to make the difference between 5D and 4D quantities, and the fifth dimension has periodicity $y\sim y+2\pi$ and is compactified on a line interval, obtained by a $Z_2$ orbifold of the circle, $S^1/Z_2$, by modding out the parity $y\to -y$.  
Thus, $R(\hat t\,)$ is the radius of the circle. The constants $H_I$ and $R_0$ are parameters corresponding to the expansion rate and the (initial) value of the radius at $\hat{t}=0$, so that $\hat{a}_0=1$.\footnote{Note the difference in the normalisation of the scale factor from the usual one, $a_t=1$ which is convenient when starting with 5D inflation but does not alter the final result, as we show in Appendix~\ref{sec:lambda_t}.}
The metric \eqref{dS5} is a solution of the 5D Einstein equations with a positive cosmological constant $\Lambda_5^I$:
\begin{align}
\label{S5}
    S_5=\int [d^4\vec{x}dy]\left\{\frac{M_*^3}{2}{\cal R}^{(5)}-
\Lambda_5^I  \right\}    \quad;\quad 6H_I^2M_*^3 = \Lambda_5^I\,,
\end{align}
where ${\cal R}^{(5)}$ is the 5D scalar curvature and $M_*$ the 5D reduced Planck mass. Validity of the effective field theory implies that both $H_I$ and the initial compactification scale $1/R_0$ are smaller than $M_*$. The brackets denote a measure transforming as a density under diffeomorphisms. $\Lambda_5^I$ can be thought of as the value of a 5D inflaton potential around an approximate flat region inducing 5D slow-roll inflation.

It is useful to describe 5D inflation in terms of a 4D background~\cite{Anchordoqui:2023etp} solving the equations of motion of an effective action containing a scalar (the radion) besides gravity, obtained upon integration over $y$ and rescaling the 4D metric to the Einstein frame:
\begin{align}
\label{S4}
ds^2 &=\frac{R_\perp}{R(t)}\left\{-dt^2+a(t)^2d\vec{x}^2\right\}+R(t)^2dy^2\nonumber\\[5pt]
 S_4 &=\int [d^4\vec{x}]\left\{\frac{{M}_p^2}{2}\left({\cal R}^{(4)}-\frac{3}{2}\left(\frac{\partial R}{R}\right)^2\right)-\frac{\pi R_\perp^2}{R}\Lambda_5^I\right\}\,,
\end{align}
where $R_\perp$ is the value of the radius at the end of inflation and $M_p^2=\pi R_\perp M_*^3$. In \eqref{S4}, we have omitted the massive spin-2 KK excitations of the 4D graviton which have vanishing background, as well as the spin-1 graviphoton 0-mode which is projected out by the $Z_2$ parity. Defining the canonically normalised radion scalar $R=R_\perp e^{\sqrt{2/3}{\sigma/M_p}}$, the action reads:
\begin{align}
\label{S4rad}
    S_4=\int [d^4\vec{x}]\left\{\frac{{M}_p^2}{2}{\cal R}^{(4)}-\frac{1}{2}(\partial\sigma)^2
    -\pi R_\perp\Lambda_5^Ie^{-\sqrt{\frac{2}{3}}{\frac{\sigma}{M_p}}}\right\}
    \quad;\quad R=R_\perp e^{\sqrt{\frac{2}{3}}{\frac{\sigma}{M_p}}}\,.
\end{align}
One thus have the usual Einstein equations coupled to a scalar field $\sigma$ with an exponential potential $V=\pi R_\perp\Lambda_5^Ie^{-\sqrt{2\over 3}{\sigma\over M_p}}$ for a Friedmann-Robertson-Walker (FRW) metric $ds_4^2=-dt^2+a^2(t)d\vec{x}^2$ with scale factor $a(t)$:
\begin{align}
\label{S4EoM}
3h^2 M_p^2&={1\over 2}\dot{\sigma}^2+V\quad;\quad h={\dot{a}\over a}\\ \nonumber
2\dot{h} M_p^2&=-\dot{\sigma}^2\,,
\end{align}
where dot denotes time differentiation. One can easily find the solution:\footnote{This solution, although exact, is not the most general one for an exponential potential. It is however the only solution with the maximum number of 15 isometries, corresponding to the symmetry group of dS$_5$, as it is obvious in the equivalent 5D action \eqref{S5}.}
\begin{align}
\label{4Dsolution}
a=a_e \left(\frac{H_It}{2}\right)^3\quad;\quad \sigma=\sqrt{6}M_p\ln\frac{H_It}{2}\quad;\quad R=R_\perp\left(\frac{H_It}{2}\right)^2\,,
\end{align}
where $H_I$ is given in \eqref{S5} and $a_e$ is the scale factor at the end of inflation $t_e=2/H_I$. It describes a power low $t^3$ inflation for the metric with an expanding radius as $t^2$. Inflation starts when the radius has initial value $R_0$ at $t_0=(R_0/R_\perp)^{1/2}t_e$. The scale factor was then 
\begin{align}
\label{a0}
t_0=(R_0/R_\perp)^{1/2}t_e:\qquad a_0=a_e(R_0/R_\perp)^{3/2}\,. 
\end{align}
Thus, $N$ e-folds of the 4D universe expansion corresponds to
$\hat{N}=2N/3$ e-folds of the radius expansion obtained by 5D
inflation.\footnote{It follows that requiring $N\gtrsim60$ to solve
  the horizon problem and taking $R_0$ near the 5D gravity scale
  $M_*^{-1}$, we  obtain $R_\perp\gtrsim1~\mu\rm{m}$, remarkably
  coinciding with the dark dimension proposal. As this was one of the
  initial motivations for the present scenario, it is important to
  note that inflation appears as a power-law expansion in the Einstein
  frame. Consequently, the Hubble radius at the onset of inflation is
  much smaller than its value at the end of the inflationary phase
  $\sim H_I^{-1}$. It follows that the relevant quantity entering the
  resolution of the horizon problem is $\hat N$ rather than $N$, as we
  shall demonstrate below.}
It will be useful for later use, to introduce a parameter $\epsilon\equiv1/(R_0M_*)\lesssim1$, in term of which the number of e-folds can be written as:
\begin{align}
\label{eq:4Defolds}
    N &\equiv\ln\left(\frac{a_e}{a_0}\right)=\frac{3}{2}\ln\left(\frac{R_\perp}{R_0}\right)=\frac{3}{2}\hat{N}\,\\
\label{eq:4Defolds2}
    &=\ln\big(R_\perp M_p\big)+\frac{1}{2}\ln\left(\frac{\epsilon^3}{\pi}\right).
\end{align}
 For later use, we also give the identity:
\begin{align}
\label{epsilonId}
\pi R_\perp M_* = (\pi R_\perp M_p)^{2/3}\quad;\quad R_\perp M_p=\frac{1}{\pi}\left(\frac{\pi}{\epsilon}\right)^{3/2}e^{N}\,.
\end{align}

Note that in the 4D theory \eqref{S4rad}, the presence of the radion background can be described as a fluid with energy density $\rho_\sigma$ and pressure $p_\sigma$ given by:
\begin{align}
    \rho_\sigma &=\frac{\dot{\sigma}^2}{2}+{V\over M_p^2}\\ \nonumber
    p_\sigma &=\frac{\dot{\sigma}^2}{2}-{V\over M_p^2}\,,
\end{align}
as can be seen from the equations of motion \eqref{S4EoM}. Using the background solution for $\sigma$ in \eqref{4Dsolution}, one obtains an equation of state $p=w\rho$ with $w=-7/9$ expected from the $t^3$ behaviour of the scale factor.
Thus, homogeneous exponential inflation in 5D can be described in terms of a 4D background for the scale factor and the radion given in \eqref{4Dsolution}, as a power law inflation. The above analysis can also be done in the Jordan frame, as seen by the 3-brane, and is presented in the Appendix~\ref{sec:B} for clarity. All results of the following sections which are obtained in the 4D Einstein frame can also be obtained in the Jordan frame in a straight forward way.

As was shown in \cite{Anchordoqui:2023etp, Antoniadis:2023sya}, despite the 4D power-low inflation background, the power spectrum of primordial fluctuations can still be scale invariant upon summation over the KK excitations of the 5D inflaton field. Their contribution becomes important when the physical 3D space momenta $k/a_e$ are larger than the compactification scale $1/R_\perp$, corresponding to small wavelengths, less than around a micron. In~\cite{Anchordoqui:2023etp,Anchordoqui:2024amx}, it was assumed that $a_0=R_0/R_\perp$ which leads to $a_e=(R_\perp/R_0)^{1/2}$ according to \eqref{a0}, creating a discontinuity in the matching of the scale factor extrapolation from the present to the past and the end of 5D inflation. Continuity requires that $a_e$ in the solution \eqref{4Dsolution} is the value of the scale factor obtained by the past history of the observable universe up to the end of inflation where $a_e=R_\perp/R_0$ and thus $a_0=(R_0/R_\perp)^{1/2}$, consistently with the rescaling of the metric in \eqref{S4}. 

We also mention the possibility of a `short' period of 4D inflation that could be driven by the radion due to its stabilization potential. Indeed after the end of 5D inflation when the inflaton is falling to the minimum of its potential, one obtains a runaway $1/R$ 4D radion potential induced by the 5D vacuum energy $\Lambda_5$ which is much different than $\Lambda_5^I$ in \eqref{S5}. Here we assume for simplicity that $\Lambda_5$ is positive. The radion can be stabilized by taking into account two additional contributions to its potential~\cite{Anchordoqui:2023etp}:\footnote{Here, we neglect the Casimir energy contribution which becomes important at a much lower scale.}
\begin{align}
\label{Vstab}
V_\text{stab}=\left(\frac{R_\perp}{R}\right)^2\hat{V}_\text{stab}\quad;\quad \hat{V}_\text{stab}=\pi R\Lambda_5+T_4+\pi\frac{K}{R}\,,
\end{align}
where the second contribution may arise from localised 3-branes and orientifolds with total tension $T_4$ and the third contribution arises for instance from kinetic gradients of bulk 5D fields~\cite{Arkani-Hamed:1999lsd} with $K$ the corresponding `flux'. Requiring (approximate) vanishing vacuum energy, $V_\text{stab}$ has a minimum for $T_4$ negative at:
\begin{align}
\label{VstabMin}
R_\perp=\left(\frac{K}{\Lambda_5}\right)^{1/2}\quad;\quad T_ 4=-2\pi\left(K\Lambda_5\right)^{1/2}
\quad;\quad m^2_\sigma=-\frac{2}{3}\frac{T_4}{M_p^2}
\end{align}
and a maximum at $R_\text{max}=3R_\perp$. It follows that $T_4$ is negative and all three terms of the potential are of the same order taking a maximum value of order $\sim (10\,\text{TeV})^4$ corresponding to the natural value of $K\sim M_*^3$. The radion mass is therefore of order of the compactification scale $(10\,\text{TeV})^2/M_p\sim{\cal O}(\text{eV})$.

In fact, it has been argued that the radion mass should be bounded from above by the compactification scale, up to an order 1 coefficient~\footnote{Nima Arkani-Hamed, private communication.} with a proposed value $2\sqrt{3}$ based on numerical evidence~\cite{Mirbabayi:2026saz}. Using the potential~\eqref{Vstab} with the minimisation~\eqref{VstabMin} and imposing the Higuchi bound at the dS maximum:
\begin{align}
\label{VstabMax}
\frac{1}{(3R_\perp)^2}\ge 2H^2=\frac{2V_\text{stab}}{3M_p^2}=-\frac{4}{81}\frac{T_4}{M_p^2}\,,
\end{align}
one obtains 
\begin{align}
\label{VstabHig}
\frac{1}{R_\perp^2}\ge -\frac{4}{9}\frac{T_4}{M_p^2}\,,
\end{align}
implying that at the minimum
\begin{align}
\label{radionbound}
m_\sigma\le \sqrt{\frac{3}{2}}m_\text{KK}\,,
\end{align}
which is slightly weaker from $\frac{2}{\sqrt{3}}m_\text{KK}$ proposed in~\cite{Mirbabayi:2026saz}.

Obviously the stabilising terms of the potential are irrelevant during inflation where the first term in \eqref{Vstab} contains $\Lambda_5^I$. However, after the end of inflation $\Lambda_5^I$ drops near $\Lambda_5$ while the radius is around the micron and its velocity is towards growing the radius $R$. Thus, the end of 5D inflation induces initial conditions for the radion in the 4D potential \eqref{Vstab}. It is plausible that the stabilization dynamics proceeds via a period of 4D inflation~\cite{Dudas:2010gi} whose analysis goes beyond the scope of this paper. 

\section{Experimental constraints on the scale of 5D inflation}
\label{sec:3}

The power spectrum of primordial scalar perturbations at 0-th order in the slow-roll approximation can be computed by considering a massless 5D inflaton around the dS$_5$ background \eqref{dS5}, or equivalently the sum of KK excitations on top of the 4D inflaton around the  background \eqref{S4EoM}. Both ways lead to the same result with a change of behaviour from the result of 4D $t^3$-inflation at  distances larger than the compactification radius with vanishing spectral index to the (5D) scale invariant spectrum at shorter distances due to the contribution of the inflaton KK tower~\cite{Anchordoqui:2023etp}.

To compute the power spectrum of cosmological fluctuations in the slow-roll approximation, it is convenient to work in the conformal time of the 5D metric \eqref{dS5}, by the transformation $\tau = - e^{-H_I \hat t}/H_I$, for which
$\hat a(\tau) = -1/(H_I\tau)$ and
$R(\tau) = -R_0/(H_I\tau)$, yielding
\begin{equation}
  ds^2 = \hat a^2 (\tau) (-d\tau^2 + dx^2 + R_0^2 \ dy^2) \, .
\label{conformal}
\end{equation}  
We then perturb the metric (\ref{conformal}) and the 5D inflaton around the
time-dependent background. As shown elsewhere~\cite{Antoniadis:2023sya,Antoniadis:2025pet}, the angular power spectrum of scalar density
perturbations can be written in terms of slow-roll parameters and is
given by\footnote{There is a difference in the overall normalisation compared with the
  expressions of~\cite{Antoniadis:2023sya,Antoniadis:2025pet} as to the mass units used is
$M_{p,0}^2\equiv 2\pi R_0{M_*}^3$, where $M_{p,0}$ is the reduced Planck
mass at the beginning of inflation. {Note that this relation holds for a circular extra dimension. In the case of an interval, the factor of $2$ is absent, since the volume is simply $\pi R_0$. However, the KK summation then acquires an additional factor of $1/2$, as only even modes are present, so that the final result remains unchanged.}}
\begin{align}\label{eq:pre_37}
    P_\mathcal{R} &\underset{R_0k \ll 1}{\simeq} \frac{H_I^3}{3\pi^4\varepsilon R_0kM_*^3}\biggl[\biggl(\frac{k}{\hat{a}H_I}\biggr)^{2\delta - 5\varepsilon}+\frac{\varepsilon}{3}\biggl(\frac{k}{\hat{a}H_I}\biggr)^{-3\varepsilon}\biggr],\\
    \label{eq:pre_47}
    P_\mathcal{R} &\underset{R_0k \gg 1}{\simeq} \frac{H_I^3}{6\pi^3\varepsilon M_*^3}\biggl[\biggl(\frac{k}{\hat{a}H_I}\biggr)^{2\delta - 5\varepsilon}+\frac{5\varepsilon}{24}\biggl(\frac{k}{\hat{a}H_I}\biggr)^{- 3\varepsilon}\biggr],
\end{align}
whereas the power spectrum of the tensor modes is found to be
\begin{eqnarray}\label{eq:pre_48}
    P_\mathcal{T} \underset{R_0k \ll 1}{\simeq} \frac{8H_I^3}{\pi^4 R_0kM_*^3}\biggl(\frac{k}{\hat{a}H_I}\biggr)^{- 3\varepsilon}
    \quad\text{and}\quad
    P_\mathcal{T} \underset{R_0k \gg 1}{\simeq}
  \frac{4H_I^3}{\pi^3M_*^3}\biggl(\frac{k}{\hat{a}H_I}\biggr)^{-
  3\varepsilon} \,,
\end{eqnarray}
where slow-roll parameters are defined by the Hubble flow
$\varepsilon = - \hat{\dot H}_I/H_I^2\equiv\varepsilon_1$ and
$\varepsilon_{n+1}={\hat{\dot\varepsilon}_n/(H_I\varepsilon_n)}$ with $\delta
= \varepsilon-\varepsilon_2/2$, where a hatted dot denotes a derivative with respect to $\hat{t}$. The ratio $r$ of the tensor-to-scalar perturbations is then $r=24\varepsilon$. 

The amplitude of the scalar power spectrum, in the region observed by
the {\it Planck} satellite  $R_0k\gg1$,  can therefore be written as:
\begin{align}
    A_s=\frac{4H_I^3}{r\pi^3M_*^3} \, ,
\label{As}
\end{align}
at lowest order in the slow-roll parameters. Note that \eqref{As} coincides with the result of 4D inflation $2H_I^2/(r\pi^2M_p^2)$ times the number of KK-excitations of the inflaton lighter than $H_I$ that contribute to density perturbations, namely $(H_IR_\perp)$.
Using the measured amplitude of adiabatic scalar perturbations $A_s\sim2.1 \times 10^{-9}$~\cite{Planck:2018vyg} and the 95\%
CL upper limit $r < 0.032$ (derived using a combination of BICEP/Keck 2018 and Planck data)~\cite{BICEP:2021xfz,Tristram:2021tvh}
we obtain an upper limit on 5D Hubble parameter during inflation:
\begin{align}
    \frac{H_I}{M_*} \lesssim 10^{-3} \,,
\label{upperHI/M*}
\end{align}
leading to
\begin{align}
    \frac{H_I}{\text{GeV}}\lesssim 5.8\times10^{5}\left(\frac{R_\perp}{\mu\text{m}}\right)^{-1/3}.
\label{upperHI}
\end{align}
The associated energy scale of inflation $M_I \sim V_I^{1/5}$ with $V_I$ the inflaton potential, linked to $H_I$ through
\begin{equation}
  6H_I^2= \frac{M_I^5}{M_*^3}
  \label{link}
\end{equation}
  is then
\begin{align}
    \frac{M_I}{\text{GeV}}\lesssim 6.0\times10^{7}\left(\frac{R_\perp}{\mu\text{m}}\right)^{-5/3}.
\end{align}
Note the the upper bound on $M_I$ is just an order of magnitude below the 5D Planck scale, which is also needed for the validity of the effective field theory of inflation, in contrast to the case of 4D inflation where a much larger hierarchy is needed with respect to the 4D Planck mass in order to obtain the observed magnitude of the power spectrum.

On a separate track, for uniform 5D inflation ($d=1$), consistency with the normalcy
temperature also requires reheating to occur at below $T_N$ given in \eqref{eq:normalcytemp}. We therefore consider a phase giving $\Delta N$ e-folds between the end of 5D inflation and the beginning of oscillations of the inflaton. 
Indeed, we adhere to the standard approach established
in~\cite{Turner:1983he} for the classical evolution of a scalar field in 
a homogeneous cosmology at the end of inflation. We are interested in the oscillations of $\phi$ about some local (or
global) minimum of  the potential $V(\phi)$. We assume that the
frequency of these oscillations, $\omega = \dot \phi/\phi$, is always much greater than
the expansion rate $h = \dot a/a$,
where $a$ is the FRW scale factor. In this limit, if the leading term in
$V(\phi)$ is $\phi^n$, the energy density of the scalar-field oscillations
red-shifts away as $a^{-6n/(n+2)}$. For $n=2$,
the energy density of the scalar-field oscillations behaves like
non-relativistic matter, whereas for $n =4$ it behaves like
relativistic matter. We reserve a detailed discussion of the model for the next section. For a general equation of state
parameter $w = \langle p \rangle/\rho$, the universe would undergo e-folds of expansion,
\begin{align}
    e^{-N_w}\sim\left(\frac{h_f}{h_i}\right)^{\frac{2}{3(1+w)}} \, ,
\end{align}
between initial and final Hubble parameters $h_i$ and $h_f$, where
\begin{equation}
\langle p \rangle = w\, \rho \,,
\end{equation}
with $\rho = \frac{1}{2} \dot \phi^2 + V(\phi)$, $p
= \frac{1}{2} \dot \phi^2 -V(\phi)$, $w = 0 \ (1/3)$
for $n=2 \ (4)$, and where the brackets indicate average over one oscillation period. Altogether, considering an RD phase after reheating one obtains the hubble parameter at reheating to be
\begin{equation}
   h_r=\sqrt{\frac{\pi^2g_*}{90}}\frac{T_r^2}{M_p},
\end{equation}
and considering that the oscillations  start when the Hubble parameter is equal to the inflaton mass, $h_{os}\sim m$, the number of e-folds during the oscillation period is given by
\begin{align}\label{Nosc}
    e^{N_{os}}=\left(\frac{h_r}{h_{os}}\right)^{-\frac{2}{3(1+w)}}=\left(\sqrt{\frac{\pi^2g_*}{90}}\frac{T_r^2}{mM_p}\right)^{-\frac{2}{3(1+w)}}.
\end{align}
We note that requiring the number of e-folds $N_{os}$ to be positive imposes the constraint
\begin{align}
\frac{m}{\text{GeV}}\geq1.4\times10^{-19}\sqrt{g_*}\left(\frac{T_r}{\text{GeV}}\right)^2
\end{align}
which is easily satisfied.

Now, at the end of 5D inflation, scale invariance is valid for all
scales below a transition wavelength  $\lambda < \lambda_t$, where the transition in the behaviour of power spectrum takes place. Equivalently, scale invariance is valid
for all co-moving momenta $k$, such that $\pi kR_0 >1$. Considering the phases that we want, inflation from $t_0$ to $t_e$, then a phase parametrised by $\Delta N$ e-folds from $t_e$ to $t_{os}$, the oscillations of the inflaton from $t_{os}$ to $t_r$, then usual cosmology from the reheating to today, namely from $t_r$ to $t_t$, we obtain
\begin{equation}\label{eq:transition}
  \pi kR_0 = \frac{2 \pi^2 a_t}{\lambda_t} \frac{R_\perp}{a_e}  = 2\pi^2 \frac{R_\perp}{\lambda_t}\frac{a_t}{a_r}\frac{a_r}{a_{os}}\frac{a_{os}}{a_e}= 2\pi^2 \frac{R_\perp}{\lambda_t}e^{N_r+N_{os}+\Delta N}\,,
\end{equation} 
giving the transition wavelength today $ \lambda_t=2\pi^2 R_\perp e^{N_r+N_{os}+\Delta N}$ where the number of e-folds since reheating is found to be~\cite{Kreling,German:2020kdp}
\begin{align}
    e^{N_r}=e^{30}\left(\frac{T_r}{\text{GeV}}\right).
\end{align}
As already mentioned in the Introduction, the transition wavelength today should be around $\lambda_t\simgt 27.3\,\text{Gpc}$. This value is derived in Appendix~\ref{sec:lambda_t} from the analysis \cite{Petretti:2024mjy}. Thus, scale invariance of the power spectrum requires
\begin{align}\label{eq:SI}
    e^{\Delta N}\geq1.0\times10^{6}\,g_*^{1/3}\left(\frac{T_r}{\text{GeV}}\right)^{1/3}\left(\frac{R_\perp}{\mu\text{m}}\right)^{-1}\left(\frac{m}{\text{GeV}}\right)^{-2/3}
\end{align}
where we considered oscillations around a quadratic potential, namely $w=0$ in \eqref{Nosc}.

Finally, the observed homogeneity of the CMB provides an empirical requirement that inflation lasted long enough to solve the horizon problem. This requires that the comoving size of the observable universe today is smaller than the comoving Hubble radius at the beginning of inflation. Namely we want:
    \begin{align}\label{horizoncon5D}
        \frac{D_{\text{LSS}}}{a_t}\lesssim\frac{1}{a_0h_0}\,,
    \end{align}
where $D_{\rm LSS}$ is defined in Appendix~\ref{sec:lambda_t} as the proper distance to the Last Scattering Surface (LSS).
Considering the phases that we introduced, the above condition becomes\footnote{As $3N/2=\hat{N}$, one would have obtain the same result by working in 5D, or in the Jordan frame, where the Hubble parameter is constant during inflation.}
    \begin{align}\label{horizoncon2}
D_{\text{LSS}}\lesssim\frac{a_t}{a_r}\frac{a_r}{a_{os}}\frac{a_{os}}{a_e}\frac{a_e}{a_0}\frac{1}{h_0}=e^{N_r+N_{os}+\Delta N+\frac{2}{3}N  }\frac{2}{3H_I}\,,
    \end{align}
where we used $h_0=3/t_0=(R_\perp/R_0)^{1/2}(3/t_e)$ and $t_e=2/H_I$ followed from the solution \eqref{S4EoM}. Using the expressions for the e-folds of the different phases, we finally obtain
\begin{align}\label{eq:hor5D}
    e^{\Delta N}\geq 2.2\times 10^{-2}\,\epsilon^{-1}\, g_*^{1/3}\left(\frac{T_r}{\text{GeV}}\right)^{1/3}\left(\frac{R_\perp}{\mu\text{m}}\right)^{-2/3}\left(\frac{H_I}{\text{GeV}}\right)\left(\frac{m}{\text{GeV}}\right)^{-2/3}.
\end{align}

\section{Low-Temperature Reheating with Suppression of Bulk Modes}
\label{sec:4}

5D inflation guarantees that bulk and brane are empty at the end of inflation while SM particles on the brane and gravitons are created by inflaton decays. Homogeneity implies that oscillations are constant in space and thus only the 4D zero-mode of the inflaton oscillates and creates particles.
Our starting point is that advanced in the previous section; namely, a 5D scalar field of mass $m$ oscillating on a
quadratic potential $V(\phi) = \frac{1}{2} m^2 \phi^2$. 
Note that $\omega = m$ and we further assume that the frequency of these
oscillations is always greater than the expansion rate, i.e., $m \gg
h$. As $\phi$ oscillates, the energy density $\rho$ becomes a slowly
varying function of time, decreasing on a time characterised by
$1/h$. However, $(\rho + p) = \dot \phi
^2$ varies rapidly, changing on a time scale characterised by
$\omega^{-1} \ll h^{-1}$; recall that for a quadratic potential $(n=2)$, $\langle p \rangle =
0$. When the 5D field oscillates, it acquires a time-dependent and space-independent background which acts as a time-dependent mass for the brane fields, say $\chi$, that
couple to the inflaton, yielding particle
creation of $\chi$ quanta~\cite{Turner:1983he}. This particle creation can be computed
classically from the time-dependent mass background, or equivalently by
computation of the widths, because the tree-level decay width
corresponds to the imaginary part of the one-loop self-energy diagram~\cite{Kofman:1994rk,Kofman:1997yn}.

Next, consider a 4D coupling between the inflaton and brane matter fields
($\phi \bar \chi \chi$), where, as previously noted, $\phi$ can be taken as a damping oscillating time-dependent background with frequency $m$ and amplitude decaying with time as $e^{(3h+\Gamma)t/2}$, 
where $\Gamma$ is the decay width~\cite{Kofman:1994rk,Kofman:1997yn}. For this setup, the particle creation is given by a rate
$\Gamma$ 
that can be obtained by the standard formula of a scalar of mass $m$ decaying for instance into 2 massless
fermions $\chi$. Now, we take the 5D Lagrangian with the coupling
multiplied by a delta function of $y$ and an oscillating
time-dependent background for $\phi$ with frequency $m$. 
As we stressed above, it is important to remind the reader that the time-dependent background
describes only the zero mode of the scalar field $\phi_0$.

The 4D action is obtained upon $y$-integration which removes
the delta function, leading to an interaction $\zeta\,\!\phi_0\,\! \bar \chi \chi$,
where $\zeta$ is a dimensionless parameter that describes the
coupling of the zero mode of the inflaton to the brane fields $\chi$. 
The partial decay width to the brane fermions $\chi$ can then be parametrised by 
\begin{equation}
  \Gamma^\phi_{\chi\chi} = \frac{\zeta^2}{8 \pi} m \, ,
\end{equation}  
where we considered the decay into fermions for the purpose of illustration. 
The arguments in the following do not depend on the particular expression of the total decay width into the brane $\Gamma^\phi_{\rm SM}$, which is as a sum of partial widths having different expressions depending on the spin of the particles and their interaction with the 5D inflaton and it is therefore model dependent.
Indeed, assuming that $\Gamma^\phi_{\rm SM}$ is the dominant decay of the inflaton, it can be determined by the reheating
temperature, which is given by
\begin{equation}
  T_r^2 \sim \Gamma^\phi_{\rm SM} M_p \, .
\end{equation}  

On the other hand, the decay rate in the bulk can be computed from the partial width of
the inflaton zero mode into gravitons
\begin{equation}
  \Gamma^{\phi}_{\rm grav} \sim \frac{m^3}{M_p^2} \ \delta n \,
\end{equation}
where $\delta n$ is the multiplicity of the KK modes that are allowed from
the decay. By demanding that $\Gamma^{\phi}_{\rm grav} \ll
\Gamma^\phi_{\rm SM}$ we obtain
\begin{equation}
  m \ll \left(\frac{T_r^2 \ M_p}{\delta n}\right)^{1/3} \, .
\label{mbound}
\end{equation}
For a multiplicity $\delta n \sim T_r R_\perp$, we have 
\begin{align}\label{eq:consmass}
    \frac{m}{\text{TeV}}\ll\left(\frac{T_r}{\text{GeV}}\right)^{1/3}\left(\frac{R_\perp}{\mu\text{m}}\right)^{-1/3}.
\end{align}
This is valid when $m\geq T_r$, otherwise $\delta n \sim mR_\perp$ and \eqref{mbound} is automatically satisfied.

The dark dimension assembles a colosseum for dark matter
contenders. For example, if the reheating temperature equals the
normalcy temperature massive spin-2 KK
excitations of the graviton provide a natural dark matter
candidate~\cite{Gonzalo:2022jac}. This is because the KK modes are non-relativistic and so
their energy density red-shifts away like $a^{-3}$. To align with
astrophysical and cosmological data one must assume that the dark
dimension lacks isometries. Indeed, if this were the case, the KK momentum of the graviton tower
would not be conserved, allowing a given KK mode of the tower to decay
into final states that include other, lighter KK excitations. These 
dark-to-dark decays provide a specific realisation of the dynamical
dark matter framework~\cite{Dienes:2011ja}.  The evolution of the
graviton tower suppresses the KK partial decay width onto SM fields to successfully accommodate astrophysical and cosmological observations~\cite{Law-Smith:2023czn, Obied:2023clp}.

If the reheating
temperature is smaller than the normalcy temperature, there
is room for other dark matter contenders. Naturally, primordial black holes with Schwarzschild radius smaller than a
micron provide one promising
candidate to complement the observed dark matter
density~\cite{Anchordoqui:2022txe,Anchordoqui:2024dxu,Anchordoqui:2025xug}. These
5D black holes are effectively
bigger, colder, and longer-lived than standard 4D black holes of the same mass. Furthermore, they have
suppressed Hawking radiation on the SM-brane and so an all-dark-matter
interpretation in terms of 5D primordial black holes becomes feasible.

\section{Constraints on the scale of 5D inflation and cosmological history below}\label{history}
We now study the cosmological evolution between the end of inflation and the mass of the inflaton, where oscillations around the minimum start; this is associated to the number of e-folds $\Delta N$ used as a parameter in our analysis above. The constraints are
\eqref{upperHI} from the power spectrum, \eqref{eq:SI} from scale invariance and \eqref{eq:hor5D} from the horizon problem, giving a lower bound on $\Delta N$ as
\begin{align}\begin{split}
    e^{\Delta N}\geq g_*^{1/3}&\left(\frac{T_r}{\text{GeV}}\right)^{1/3}\left(\frac{R_\perp}{\mu\text{m}}\right)^{-2/3}\left(\frac{m}{\text{GeV}}\right)^{-2/3}\\
    &\times\max\left\{1.0\times10^{6}\left(\frac{R_\perp}{\mu\text{m}}\right)^{-1/3},\,2.2\times10^{-2}\,\epsilon^{-1}\left(\frac{H_I}{\text{GeV}}\right)\right\}.\end{split}
\end{align}
In order to estimate the minimum value of the common factor, one can use \eqref{eq:consmass} to write 
\begin{align}
    g_*^{1/3}&\left(\frac{T_r}{\text{GeV}}\right)^{1/3}\left(\frac{R_\perp}{\mu\text{m}}\right)^{-2/3}\left(\frac{m}{\text{GeV}}\right)^{-2/3}\gg10^{-2}g_*^{1/3}\left(\frac{T_r}{\text{GeV}}\right)^{1/9}\left(\frac{R_\perp}{\mu\text{m}}\right)^{-4/9}
\end{align}
and using $T_r\geq T_\text{BBN}\sim5\,\text{MeV}$ and $R_\perp\leq40\,\mu\text{m}$ we finally obtain
\begin{align}
    g_*^{1/3}&\left(\frac{T_r}{\text{GeV}}\right)^{1/3}\left(\frac{R_\perp}{\mu\text{m}}\right)^{-2/3}\left(\frac{m}{\text{GeV}}\right)^{-2/3}\gg10^{-3}.
\end{align}
Considering $\epsilon\leq1$, we can extremise the condition as
\begin{align}
    e^{\Delta N}\gg\max\left\{1.0\times10^{3}\,,\,2.2\times\,10^{-5}\,\left(\frac{H_I}{\text{GeV}}\right)\right\}.
\end{align}
One can finally use \eqref{upperHI} to see that the second term is the smallest one, so that the condition becomes
\begin{align}
    e^{\Delta N}\gg10^3\,,
\end{align}
yielding $\Delta N>7$.

In order to obtain an intuition for the implications of this estimate, let us consider an evolution driven by a constant equation of state parameter $w$, so that the expansion during the intermediate phase after the end of inflation to the beginning of oscillations can be parametrised as 
\begin{align}
    e^{\Delta N}=\left(\frac{2m}{3H_I}\right)^{-\frac{2}{3(1+w)}}
\end{align}
where we used that $h_e=3H_I/2$. The positivity of $\Delta N$ then implies the condition 
\begin{align}
    m\leq\frac{3H_I}{2}.
\end{align}
The different phases of the cosmological evolution considered above are then summarised in Table~\ref{tab:phases1}.

\begin{table}[h]
\centering
\renewcommand{\arraystretch}{1.5} 

\begin{tabular}{r|c|c|c|l}
\cline{2-4}
\smash{\raisebox{-12pt}{beginning of inflation $\rightarrow$}}& $w_5$ & $w_4$ & $N$ & \smash{\raisebox{-12pt}{$\leftarrow$ $R_0$}}\\ 
\hhline{~===~}

\smash{\raisebox{-10pt}{end of inflation $\rightarrow$}}& $-1$ & $-7/9$ & $N$& \smash{\raisebox{-10pt}{$\leftarrow$ $R_\perp$}} \\ 
\cline{2-4}
\smash{\raisebox{-10pt}{beginning of oscillations $\rightarrow$}}&  & $w$ & $\Delta N$ &  \\
\cline{2-4}
\smash{\raisebox{-10pt}{reheating $\rightarrow$}}&  & $0$ & $N_{os}$ &  \\
\cline{2-4}
\smash{\raisebox{-10pt}{today $\rightarrow$}}&  & $\sim1/3$ & $N_r$ &  \\
\cline{2-4}

\end{tabular}
\caption{The different phases of the cosmological evolution}
\label{tab:phases1}
\end{table}
It is easy to see that consistency of the constraint
\eqref{upperHI} from the power spectrum with the scale invariance condition \eqref{eq:SI} imposes $w\lesssim0$. 
The value $w=0$ which saturates the bound seems to be marginally excluded. Indeed, setting $w=0$ actually removes the dependence of the constraint on the inflaton mass $m$ as the scale invariant condition \eqref{eq:SI} becomes
\begin{align}
   \left(\frac{H_I}{\text{GeV}}\right)\geq6.9\times10^{8}g_*^{1/2}\left(\frac{T_r}{\text{GeV}}\right)^{1/2}\left(\frac{R_\perp}{\mu\text{m}}\right)^{-3/2}
\end{align}
while the horizon problem constraint \eqref{eq:hor5D} becomes
\begin{align}
    \left(\frac{H_I}{\text{GeV}}\right)\leq2.2\times10^{5}\,\epsilon^3\, g_*^{-1}\left(\frac{T_r}{\text{GeV}}\right)^{-1}\left(\frac{R_\perp}{\mu\text{m}}\right)^{2}.
\end{align}
However, even going to extreme low values of $T_R\sim T_{\text{BBN}}\sim5\,\text{MeV}$ (corresponding to $g_*=10.74$ \cite{Husdal:2016haj}) and $R_\perp=40\,\mu\text{m}$ the scale invariance condition gives $H_I\geq6.4\times10^5\,\text{GeV}$, in contrast with the CMB constraint \eqref{upperHI}. 

\begin{figure}[h]
     \begin{subfigure}[t]{0.5\textwidth}
    \centering\includegraphics[width=0.8\linewidth]{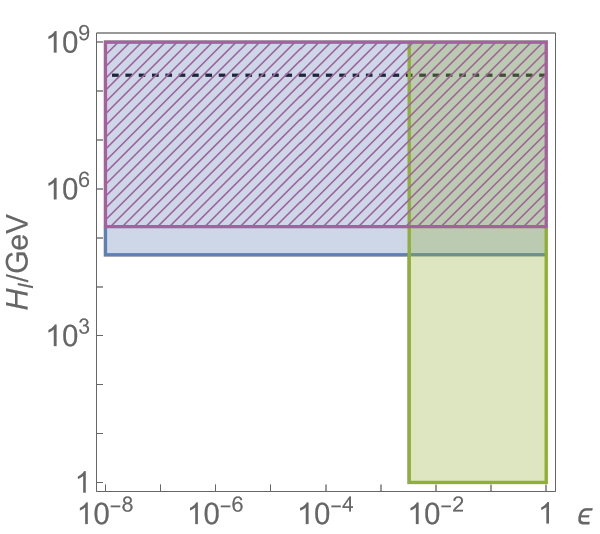}
    \captionsetup{skip=-1pt}
    \subcaption{$R_\perp\!=\!40\,\mu\text{m},\ T_r\!=\!290\,\text{MeV}\!\simeq\! T_N,\ m\!=\!1\,\text{GeV}$}
    \label{fig:1}
    \end{subfigure}
    \begin{subfigure}[t]{0.5\textwidth}
    \centering\includegraphics[width=0.8\linewidth]{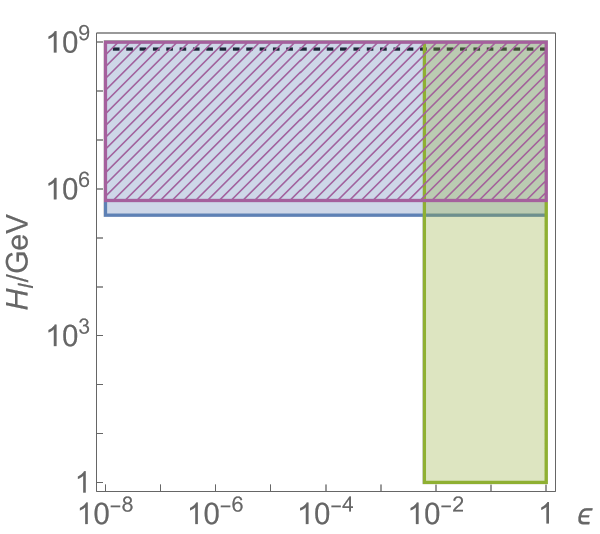}
    \captionsetup{skip=-1pt}
    \subcaption{$R_\perp\!=\!1\,\mu\text{m},\ T_r\!=\!1\,\text{GeV}\!\simeq\! T_N,\ m\!=\!1\,\text{MeV}$}
    \label{fig:2}
    \end{subfigure}
  \begin{subfigure}[t]{0.5\textwidth}
    \centering\includegraphics[width=0.8\linewidth]{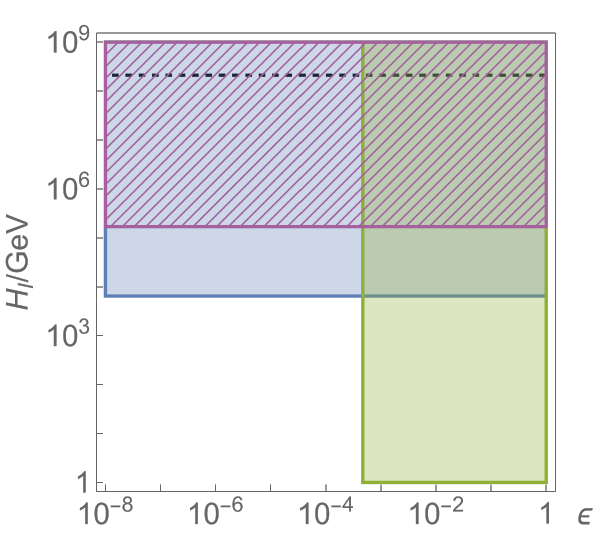}
    \captionsetup{skip=-1pt}
    \subcaption{$R_\perp\!=\!40\,\mu\text{m},\ T_r\!=\!5\,\text{MeV}\!\gtrsim\! T_{\text{BBN}},\ m\!=\!1\,\text{GeV}$}
    \label{fig:3}
    \end{subfigure}
    \begin{subfigure}[t]{0.5\textwidth}
    \centering\includegraphics[width=0.8\linewidth]{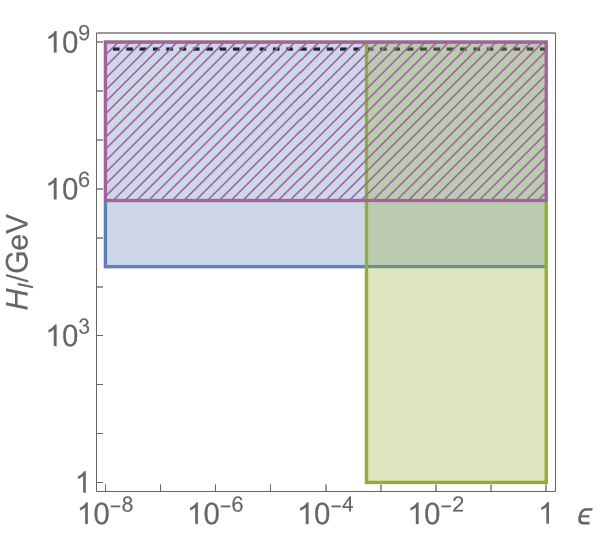}
    \captionsetup{skip=-1pt}
    \subcaption{$R_\perp\!=\!1\,\mu\text{m},\ T_r\!=\!5\,\text{MeV}\!\gtrsim\! T_{\text{BBN}},\ m\!=\!1\,\text{MeV}$}
    \label{fig:4}
    \end{subfigure}
    \begin{subfigure}[t]{0.5\textwidth}
    \centering\includegraphics[width=0.8\linewidth]{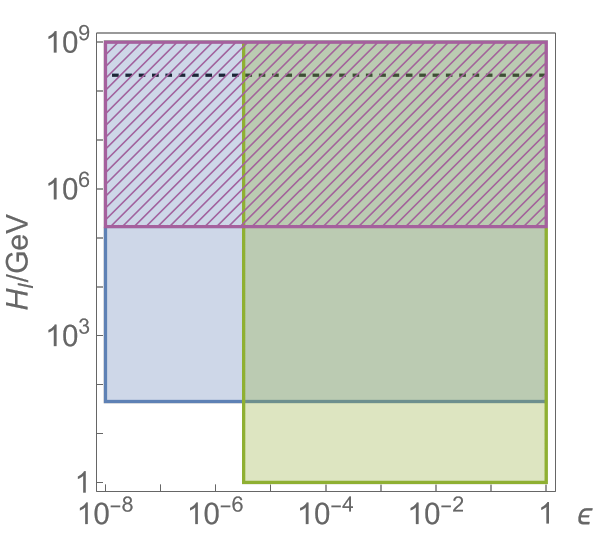}
    \captionsetup{skip=-1pt}
    \subcaption{$R_\perp\!=\!40\,\mu\text{m},\ T_r\!=\!290\,\text{MeV}\!\simeq\! T_{N},\ m\!=\!1\,\text{eV}$}
    \label{fig:5}
    \end{subfigure}
    \begin{subfigure}[t]{0.5\textwidth}
    \centering\includegraphics[width=0.8\linewidth]{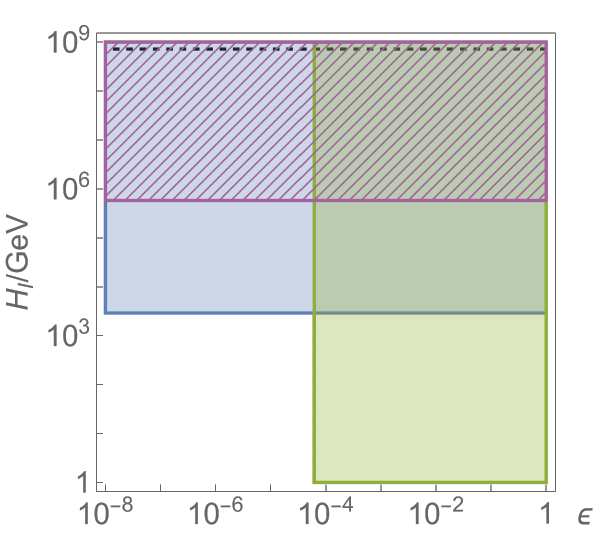}
    \captionsetup{skip=-1pt}
    \subcaption{$R_\perp\!=\!1\,\mu\text{m},\ T_r\!=\!1\,\text{GeV}\!\simeq\! T_{N},\ m\!=\!1\,\text{eV}$}
    \label{fig:6}
    \end{subfigure}
  \caption{Constraints on 5D inflation. The blue and green regions are respectively the scale invariance condition \eqref{eq:SIw=-1/3}, the horizon problem constraint \eqref{eq:horw=-1/3}. The allowed region of the parameter space is therefore the intersection of the two. The purple hashed region is excluded by the CMB observations \eqref{upperHI} and the black dashed line is the value of $M_*$. }
  \label{fig:constraints5D}
\end{figure}

Let us now consider the limiting case $w=-1/3$ of a non-accelerating expansion, corresponding to an expansion linear in time, which is of particular phenomenological interest because it yields the largest possible number of e-folds attainable during this epoch while remaining outside the regime of four-dimensional inflation. In this case, the scale invariance condition \eqref{eq:SI} becomes
\begin{align}\label{eq:SIw=-1/3}
    \left(\frac{m}{\text{GeV}}\right)^{1/3}\leq1.5\times10^{-6}g_*^{-1/3}\left(\frac{T_r}{\text{GeV}}\right)^{-1/3}\left(\frac{R_\perp}{\mu\text{m}}\right)\left(\frac{H_I}{\text{GeV}}\right)
\end{align}
while the horizon constraint \eqref{eq:hor5D} becomes independent of $H_I$:
\begin{align}\label{eq:horw=-1/3}
    \left(\frac{m}{\text{GeV}}\right)^{1/3}\leq70\,\epsilon\, g_*^{-1/3}\left(\frac{T_r}{\text{GeV}}\right)^{-1/3}\left(\frac{R_\perp}{\mu\text{m}}\right)^{2/3}.
\end{align}
They are shown in Fig.~\ref{fig:constraints5D} for different values of
the parameters.\footnote{We use $g_*=10.74,\, 63$ and $76.34$ for
  $T_r=5\,\text{MeV},\,290\,\text{MeV}$ and $1\,\text{GeV}$
  respectively \cite{Husdal:2016haj}.}  For illustration, we vary $m$ in
the range $eV < m < GeV$ and plot the allowed region of $H_I$ as a function of $\epsilon$ for various values of $T_t$ and $R_\perp$.
Finally, we note the possibility that $e^{\Delta N}$ contains a phase of 4D inflation, as discussed in Sec.~\ref{setup}. This would relax the constraints on the model parameters, thereby improving the phenomenological viability of the scenario.

\section{Two micron-size extra dimensions}
\label{d=2}

We now turn to the case of two-extra dimensions; namely a scenario of 6D uniform inflation. Collider
physics sets a lower bound on the 6D gravity scale $M_* \gtrsim 10~{\rm TeV}$~\cite{Anchordoqui:2025nmb}. This in turn places an upper limit on the compactification radius; namely, $R_\perp \lesssim 1~\mu{\rm m}$. The 4D background \eqref{4Dsolution} becomes~\cite{Anchordoqui:2023etp}:
\begin{align}
\label{4Dsolutiond=2}
a=a_e (H_It)^2\quad;\quad \sigma=2M_p\ln(H_It)\quad;\quad R=R_\perp (H_It)=R_\perp e^{-\frac{1}{2}\frac{\sigma}{M_p}}\,,
\end{align}
with 
\begin{align}
\label{a0d=2}
t_e=1/H_I\quad;\quad t_0=\left(\frac{R_0}{R_\perp}\right)t_e\quad;\quad a_0=a_e(R_0/R_\perp)^{2}\,,
\end{align}
implying that the number of e-folds of the 4D universe expansion is double than the number of the radius expansion. Thus, \eqref{eq:4Defolds} becomes
\begin{align}
\label{eq:4Defoldsd=2}
N=2\ln\left(\frac{R_\perp}{R_0}\right)=\ln(R_\perp M_p) + \ln\left(\frac{\epsilon^2}{\pi}\right)\,,
\end{align}
where we used $\pi R_\perp M_p = (\pi R_\perp M_*)^2$. Finally, continuity implies $a_e=R_\perp/R_0$ and thus $a_0=R_0/R_\perp$.

Firstly, the analysis of Sec.~\ref{sec:normalcytemp} can directly be extended in the case of two extra dimensions by replacing the multiplicity of the KK modes by $\delta n=(TR_\perp)^2$ in \eqref{eq:gammabath}. This leads to
\begin{align}\label{eq:T_N6D}
    \frac{T_N}{\text{MeV}}\simeq4\left(\frac{R_\perp}{\mu\text{m}}\right)^{-1/2}.
\end{align}
A point worth noting at this juncture is that after correcting for the
value of the Hubble constant ($h=0.67$ rather than 0.75) and
normalizing the results of~\cite{Hannestad:2001nq} to the reduced
Planck mass $M_p$ rather than the Planck mass, our calculations agree
with those in Fig.~1 of~\cite{Hannestad:2001nq}. The fundamental mass
scales $M$ (labelling the vertical axis in Fig.~1 of~\cite{Hannestad:2001nq}) and $M_*$ relate as $M^4 = 8 \pi^3 M_*^4$.

Next, the amplitude of the power spectrum of scalar density perturbations is \cite{Hirose:2025pzm} 
{\begin{equation}
    A_s=\frac{3H_I^4}{\pi^3rM_*^4}
\end{equation}
with $r=32\varepsilon$.}
This implies 
\begin{align}\label{eq:HICMB6D}
    \frac{H_I}{\text{GeV}}\lesssim 64\left(\frac{R_\perp}{\mu\text{m}}\right)^{-1/2},
\end{align}
or using $ 10H_I^2 = M_I^6/M_*^4$ one obtains  
\begin{equation}
\frac{M_I}{\text{GeV}}\leq 3.2\times10^{3}\left(\frac{R_\perp}{\mu\text{m}}\right)^{-1/2}.
  \end{equation}  
  
We turn to the constraint from the condition that the
inflaton decay into SM particles dominates over its decay on bulk
modes. Duplicating the analysis carried out in Sec.~\ref{sec:4} but taking
$d=2$, it is easily seen that the relation $\Gamma^\phi_{\rm grav} \ll
\Gamma^\phi_{\rm SM}$ also leads to the constraint (\ref{mbound}). For
$d=2$, however, we
have $\delta n \sim (T_r R_\perp)^2$ so that $T_r$ dependence drops and we obtain
\begin{equation}
  \frac{m}{\text{GeV}}\ll \left(\frac{R_\perp}{\mu\text{m}}\right)^{-2/3} \, .
\label{mmmm}
\end{equation}  
Again, this is valid when $m\ge T_r$, otherwise $\delta n \sim (mR_\perp)^2$ and \eqref{mbound} is automatically satisfied.

\begin{figure}[h]
     \begin{subfigure}[t]{0.5\textwidth}
    \centering\includegraphics[width=0.8\linewidth]{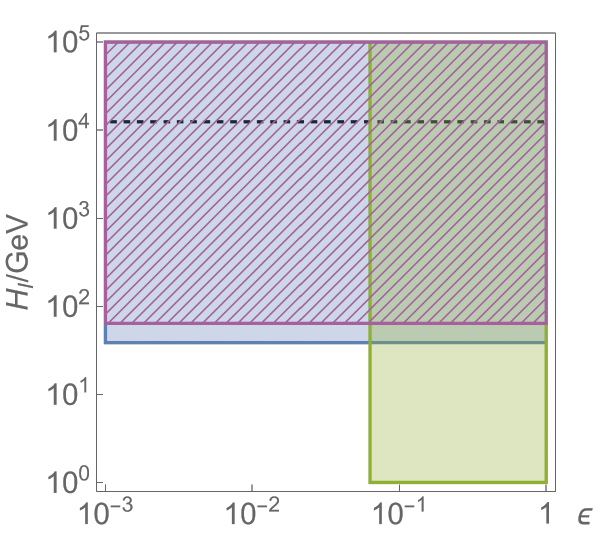}
    \captionsetup{skip=-1pt}
    \subcaption{$R_\perp\!=\!1\,\mu\text{m},\ T_r\!=\!T_N\!\simeq\!4\,\text{MeV},\ m\!=\!10\,\text{meV}$}
    \label{fig:1_6D}
    \end{subfigure}
    \begin{subfigure}[t]{0.5\textwidth}
    \centering\includegraphics[width=0.8\linewidth]{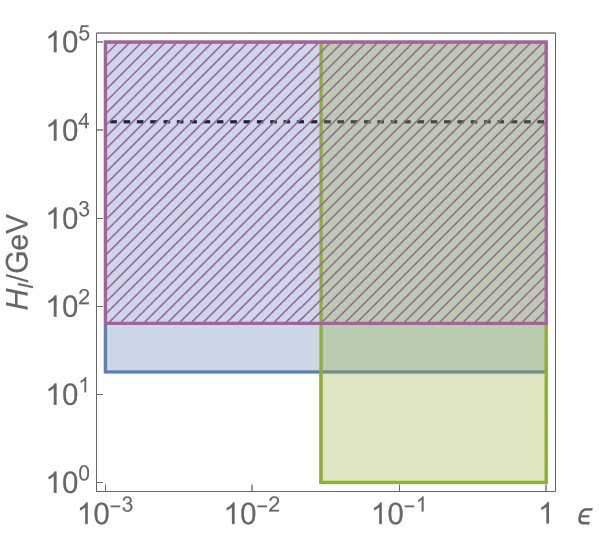}
    \captionsetup{skip=-1pt}
    \subcaption{$R_\perp\!=\!1\,\mu\text{m},\ T_r\!=\!T_N\!\simeq\!4\,\text{MeV},\ m\!=\!1\,\text{meV}$}
    \label{fig:2_6D}
    \end{subfigure}
  \begin{subfigure}[t]{0.5\textwidth}
    \centering\includegraphics[width=0.8\linewidth]{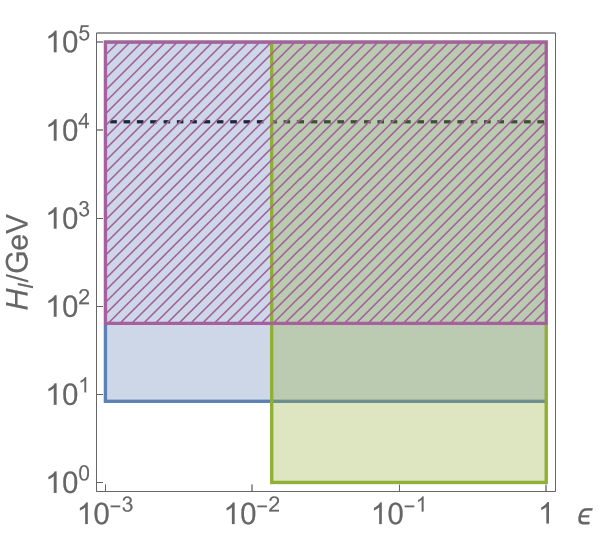}
    \captionsetup{skip=-1pt}
    \subcaption{$R_\perp\!=\!1\,\mu\text{m},\ T_r\!=\!T_N\!\simeq\!4\,\text{MeV},\ m\!=\!0.1\,\text{meV}$}
    \label{fig:3_6D}
    \end{subfigure}
    \begin{subfigure}[t]{0.5\textwidth}
    \centering\includegraphics[width=0.8\linewidth]{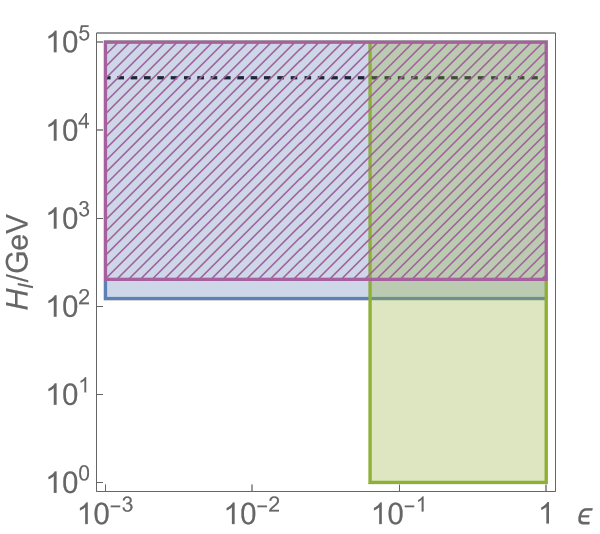}
    \captionsetup{skip=-1pt}
    \subcaption{$R_\perp\!=\!0.1\,\mu\text{m},\ T_r\!=\!T_N\!\simeq\!13\,\text{MeV},\ m\!=\!0.1\,\text{meV}$}
    \label{fig:4_6D}
    \end{subfigure}

  \caption{Constraints on 6D inflation. The blue and green regions are respectively the scale invariance condition \eqref{eq:SI6Dw=-1/3}, the horizon problem constraint \eqref{eq:hor6Dw=-1/3}. The allowed region of the parameter space is therefore the intersection of the two. The purple hashed region is excluded by the CMB observations \eqref{eq:HICMB6D} and the black dashed line is the value of $M_*$. }
  \label{fig:constraints6D}
\end{figure}

Finally, we notice that the derivation of the scale invariance condition $\eqref{eq:SI}$ only relies on the post-inflationnary evolution of the Universe and, therefore, does not depend on the number of extra dimensions. On the contrary, solving the horizon problem now requires
\begin{align}\label{eq:hor6D}
    e^{\Delta N}\geq 1.7\times 10^{3}\,\epsilon^{-1}\, g_*^{1/3}\left(\frac{T_r}{\text{GeV}}\right)^{1/3}\left(\frac{R_\perp}{\mu\text{m}}\right)^{-1/2}\left(\frac{H_I}{\text{GeV}}\right)\left(\frac{m}{\text{GeV}}\right)^{-2/3}\,,
\end{align}
which combined with the scale invariance constraint gives
\begin{align}\label{bracketconstr}
\begin{split}
    e^{\Delta N}\geq g_*^{1/3}&\left(\frac{T_r}{\text{GeV}}\right)^{1/3}\left(\frac{R_\perp}{\mu\text{m}}\right)^{-1/2}\left(\frac{m}{\text{GeV}}\right)^{-2/3}\\
    &\times\max\left\{1.0\times10^{6}\left(\frac{R_\perp}{\mu\text{m}}\right)^{-1/2},\,1.7\times10^{3}\,\epsilon^{-1}\left(\frac{H_I}{\text{GeV}}\right)\right\}\,.\end{split}
\end{align}
It follows that
\begin{align}\begin{split}
    g_*^{1/3}\left(\frac{T_r}{\text{GeV}}\right)^{1/3}\left(\frac{R_\perp}{\mu\text{m}}\right)^{-1/2}\left(\frac{m}{\text{GeV}}\right)^{-2/3}&\gg g_*^{1/3}\left(\frac{T_r}{\text{GeV}}\right)^{1/9}\left(\frac{R_\perp}{\mu\text{m}}\right)^{-1/18}\\
    &\gg0.54\,g_*^{1/3}\left(\frac{R_\perp}{\mu\text{m}}\right)^{-1/9}\\
    &\gg1
    \end{split}
\end{align}
where in the  second line we used $T_r\sim T_N$ as it is also of the order of $T_{\text{BBN}}$ and in the last one we used $R_\perp\lesssim1\,\mu\text{m}$. One can further use \eqref{eq:HICMB6D} to show that the second term in the bracket of \eqref{bracketconstr} is always the smallest so that
\begin{align}
    e^{\Delta N}\gg 10^6.
\end{align}
Again, in order to maximise $\Delta N$ without invoking a phase of 4D inflation, we introduce a phase of liear expansion
\begin{align}
    e^{\Delta N}=\left(\frac{m}{2H_I}\right)^{-\frac{2}{3(1+w)}}\qquad\qquad\text{with}\qquad\qquad w=-1/3.
\end{align}
By further looking at $T_r=T_N$ given in \eqref{eq:T_N6D}, the scale invariance condition becomes
\begin{align}\label{eq:SI6Dw=-1/3}
    \left(\frac{m}{\text{GeV}}\right)^{1/3}\leq1.2\times10^{-5}\,g_*^{-1/3}\left(\frac{R_\perp}{\mu\text{m}}\right)^{7/6}\left(\frac{H_I}{\text{GeV}}\right)
\end{align}
while the horizon problem constraint is
\begin{align}\label{eq:hor6Dw=-1/3}
    \left(\frac{m}{\text{GeV}}\right)^{1/3}\leq7.5\times10^{-3}\,\epsilon\, g_*^{-1/3}\left(\frac{R_\perp}{\mu\text{m}}\right)^{2/3}.
\end{align}
On the other hand, comparing \eqref{eq:SI6Dw=-1/3} with \eqref{eq:HICMB6D}, one can see that the inflaton mass is bounded by $m\lesssim 4.5\times 10^{-2}\,\text{eV}$. The allowed parameter space is shown in Fig.~\ref{fig:constraints6D} for  different values of the parameters.\footnote{We use $g_*=10.74$ and $10.76$ for $T_r=4$ and $13\,\text{MeV}$ respectively \cite{Husdal:2016haj}.} We emphasise again that a brief period of 4D inflation following the 5D expansion may relax these constraints.

\section{Conclusions}
\label{sec:6}

In this work, we proposed a cosmological history that connects a
period of higher dimensional inflation with standard cosmology. Higher
dimensional inflation blows up the size of compact dimensions from an
initial value $R_0$ 
to a much larger size needed to fix the present strength of gravity, while at the same time it expands our observable universe in the three non-compact dimensions to a size needed for explaining the causal horizon problem. It thus provides a minimal framework for  relating and explaining two large hierarchies in nature: the weakness of gravitational interactions and the largeness of the observable universe. 

It was also realised that higher dimensional inflation can lead to an approximate scale invariant power spectrum of primordial scalar perturbations consistent with observation if the size of extra dimensions is around the micron range. Combined with the experimental bounds on Newton's inverse square law, and particle physics and astrophysical constraints, there are only two possible cases for the number $d$ of micron-size extra dimensions $d=1,2$. The fundamental (higher dimensional) gravity scale $M_*$ in these cases lies in the region of $\sim 10^9$ GeV for $d=1$ and $\sim 10$ TeV for $d=2$. This experimental information combines nicely with the recent dark dimension proposal for addressing the smallness of the dark energy, as well as with the past framework for addressing the mass hierarchy problem in the case $d=2$.

One of the main constraints for implementing large extra dimensions in standard cosmology comes from the thermal overproduction of graviton KK modes, implying the so-called normalcy temperature $T_N$ above which the Universe should be empty of particles and bulk modes, while below $T_N$ one can have the normal standard cosmology. It turns out that the normalcy temperature is quite low, around 1 GeV for $d=1$ and 4 MeV for $d=2$, that constitutes a major constraint for making a consistent model of the Universe evolution at earlier times starting near the fundamental gravity scale and implementing inflation.
In this work, we addressed this issue and proposed a concrete cosmological history that connects higher dimensional inflation and its successful predictions with standard cosmology that starts at a reheating temperature, below or of order $T_N$. 

We first note that higher dimensional inflation guarantees that at the end of it, the Universe becomes homogeneous and empty of particles on the brane and on the bulk. Subsequently, the `long' period between the end of inflation at a scale $M_I$ and the normalcy temperature $T_N$ is described by two phases: 
\begin{itemize} 
\item Damping in time oscillations of the inflaton field around the minimum of its potential, parametrised (in the case of a quadratic approximation) by a pressure-less fluid that makes the Universe to expand along the three non-compact dimensions as matter dominated, assuming a radius stabilization at the end of inflation, for instance based on the mechanisms proposed in~\cite{Anchordoqui:2023etp} and we elaborated here. The damping of oscillations is mainly driven by the decay of the inflaton into SM particles on the brane, while its decay into bulk modes is negligible. This period starts when the expansion rate of the Universe is lower than the inflaton mass $m$. 
\item A second period from the end of 5D inflation up to the inflaton mass that as a first step we kept it unknown, keeping in the constraints as parameter of its duration as additional number of e-folds, driven by an effective cosmological fluid with an equation of state depending on a constant parameter $w$ which is constrained to be negative. A phenomenologically interesting case is $w=-1/3$ corresponding to a limiting case of non accelerating neither decelerating expansion, linear in the proper time.
\end{itemize}

We then imposed experimental constraints from {\it (i)}~the amplitude of the power spectrum of primordial fluctuations observed on the brane, leading to an upper bound on the scale of inflation $M_I$ which is an order of magnitude below $M_*$, {\it (ii)}~the transition point in the CMB spectrum where deviations from scale invariance are allowed, leading to a lower bound on the inflaton mass depending on the number of e-folds of the 2nd phase, and {\it (iii)}~the solution of the causal horizon problem, leading to a lower bound on the initial value of the compactification scale depending on the other parameters; the bound simplifies significantly in the limiting case of linear expansion where the dependence on the inflationary Hubble parameter $H_I$ drops. Combining all constraints, we determined the allowed parameter space, showing that the constraints can be satisfied with natural values of the parameters without the need of large hierarchies and tunings. In particular, the scale of 5D inflation can be an order of magnitude below the 5D Planck scale $M_*$, in contrast to the case of 4D inflation, and the initial value of the compactification scale can be within the region of $M_*$ and the Hubble scale of inflation, while the inflaton mass can be near the normalcy temperature. The main observational predictions of 5D inflation are the deviations from the simple form of scale invariance which are suppressed by the slow roll parameter that fixes the amplitude of primordial gravitational waves and can be as large as the present experimental bound. Of course, there is also the rise of the power spectrum at large angular scales which is rather difficult to detect because of the statistical fluctuations arising from cosmic variance. 

Our analysis in this work did not take into account a possible important theoretical constraint from the {\it transPlanckian censorship conjecture} (TCC)~\cite{Bedroya:2019snp}. It is supported by arguments based in the asymptotic behaviour of scalar potentials that do not allow eternal inflation and restricts significantly the number of e-folds and the scale of 4D inflation. However, when inflation occurs for some finite period in the interior of moduli space, it requires extra assumptions on both the transition to the asymptotic region, as well as on the future of the universe evolution. A dedicated analysis of TCC in the context of the dark dimension will be the subject of a separate forthcoming publication.

Finally, the problem of a concrete working mechanism of the radius stabilisation needs to be implemented consistently with the proposed cosmological history.

\section*{Acknowledgements}

We are grateful to Savas Dimopoulos for valuable discussions and collaboration at the initial stage of this work.
I.A. would like to thank Nima Arkani-Hamed, Renata Kallosh and Andrei
Linde for enlightening discussions. I.A. and L.A.A. would like to
thank the hospitality of the Center for Cosmology and Particle Physics
in NYU where part of this work was performed. The work of L.A.A. is
supported by the U.S. National Science Foundation (NSF Grant
PHY-2412679), he extends his appreciation to the Harvard Swampland
Initiative for their warm hospitality and for providing a stimulating
environment for productive discussions. The research of I.A. was supported in part by the Higher Education and Science Committee of MESCS RA (Research Project N 24RL-1C036).

\appendix

\section{Assessing the impact 
of KK decays to relic density calculations}
\label{App:T_N}

In this Appendix, we evaluate how KK decays could affect our estimate
of the normalcy temperature. We first consider the scenario of
standard large extra dimensions, as formulated
in~\cite{Arkani-Hamed:1998jmv,Antoniadis:1998ig}. Compactifying
$n$ large dimensions of radius $R_{\perp }$ allows bulk fields to be
Fourier expanded into KK modes labeled by $\vec l = (l_1, l_2, \cdots,
l_n)$. The quantized momentum, given by ${\vec p}_n^{\;2} = \vec l
\cdot \vec l/R_\perp^2$, appears to a 4D observer as a KK tower of
states with mass $m^2_{l} = \vec p_n^{\;2}$ and identical spin and
gauge quantum numbers. Conservation of extra-dimensional momentum
requires that $\vec l_{\rm i} = \vec l_{{\rm f}_1} + \vec l_{{\rm
    f}_2}$. Since the initial mass must equal the sum of the final
masses, $m_{l_{\rm i}} =\sqrt{(\vec l_{{\rm f}_1} + \vec l_{{\rm
      f}_2})^2}/ R_\perp$, there is no available (ordinary 3D) phase
space for the decay, making it kinematically forbidden. However,
confining SM fields to the brane explicitly violates
higher-dimensional translational invariance. Since transverse momentum
is no longer conserved in this setup, KK modes are allowed to decay
into SM particles.

\begin{figure}[tbh]
  \postscript{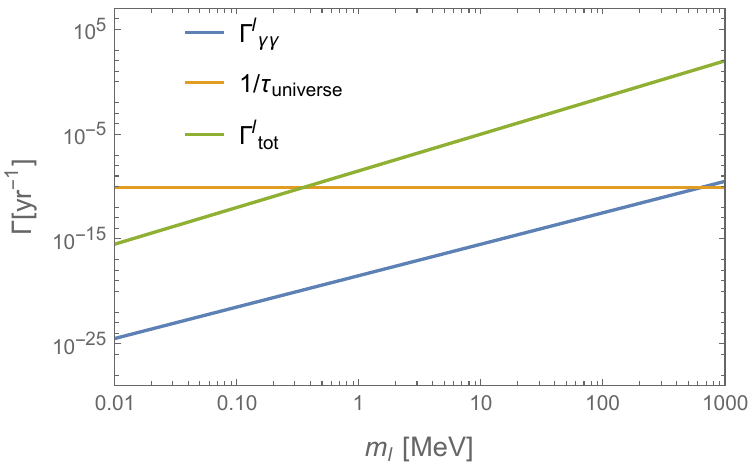}{0.75}
  \caption{Comparison of decay widths with the Hubble rate today ($\sim 1/\tau_{\rm universe}$. We have taken $\tilde \lambda = \delta = \beta =1$. \label{fig:Gammas}}
\end{figure}

In particular, the partial decay width of the radiative decay of KK
gravitons into photons is found to be
\begin{equation}
  \Gamma^{l}_{\gamma \gamma} = \frac{{\tilde \lambda}^2 \ m^3_l}{80 \pi
    M_p^2} \, ,
\label{gammagamma}
\end{equation}
where the parameter $\tilde \lambda$ measures the value of the dark graviton wave
function at the SM brane and is expected to be ${\cal
  O}(1)$~\cite{Han:1998sg,Hall:1999mk}. In Fig.~\ref{fig:Gammas} we compare the decay rate $\Gamma _{\gamma \gamma
}^{l}$ (with $\tilde \lambda =1$) to the Hubble rate today 
for relevant masses $m_{l} \lesssim 1~{\rm GeV}$. As shown in the
figure, the KK lifetime ($\sim 1/\Gamma^l_{\gamma \gamma}$) is
comparable to or exceeds the age of the universe. In fact, across the
majority of this mass range, the KK lifetime far exceeds the age of
the universe. Present relic density measurements lack the precision
needed for KK decays to affect our determination of the normalcy temperature~\cite{Arkani-Hamed:1998sfv,Hannestad:2001nq}. Conversely, KK with $m_{l} \sim 1~{\rm GeV}$ can skew
precision CMB and astrophysical data, which ultimately sets a
definitive constraint on $\tilde{\lambda }$~\cite{Law-Smith:2023czn,Obied:2023clp}.

We now assume that extra dimensions do not admit
  isometries, whereby conservation of the extra dimensional momentum is
  violated, allowing the massive KK modes
  of the graviton to decay to other lighter graviton
  modes. For $n=1$, the total
  decay width of graviton $l$ is estimated to be, 
\begin{eqnarray}
  \Gamma^{l}_{\rm tot} \sim \beta^2 \
    \delta^{3/2} \frac{m_l^{7/2}}{M_p^2 m_{\rm KK}^{1/2}} \,,
\label{Gtot}
\end{eqnarray}   
 where $m_{\rm KK} \sim 1~{\rm eV}$, $\delta$ controls the velocity of the decay products
 \begin{equation}
  v \sim \sqrt{m_{\rm KK} \ \delta /m_l}\,,
\label{velocity}
\end{equation}
and $\beta \sim 1$ is parameter that controls the strength of the intra-tower
decay amplitudes, which correlates with the amplitudes on
inhomogeneities in the dark dimension~\cite{Gonzalo:2022jac}. For $n=2$, we have
\begin{eqnarray}
  \Gamma^{\vec l}_{\rm tot} \sim \beta^2 \
    \delta^{3/2} \frac{m_l^{9/2}}{M_p^2 m_{\rm KK}^{3/2}} \, .
\label{Gtot2}
\end{eqnarray}
 We further assume that a parent particle with mass $m_l$ can only decay to
two daughter particles with masses $m_{l'}$ and $m_{l''}$ such that
$m_l = m_{l'} + m_{l''} + \epsilon$, with $\epsilon \leq 
m_{\rm KK} \delta$ and $\delta \sim 1$~\cite{Gonzalo:2022jac}. We provide justification for this
assumption below.

For $\delta \sim 1$, conservation of energy forces the momenta of
the decay products to be almost parallel and so for $n=2$, the number
of decay channels available is still effectively 1D, up to a width
that will be set by $\delta$~\cite{Anchordoqui:2025nmb}. Thus, under the assumption of $\delta
\sim 1$ the total decay width for $n=2$ is given by (\ref{Gtot})
rather than (\ref{Gtot2}).

As shown in Fig.~\ref{fig:Gammas}, for masses below 10~keV, the dark-to-dark decay
width dominating the graviton tower evolution is significantly smaller
than the Hubble rate today. Thus, despite the shift in the actual mass
of the dark matter particles, they all behave as CDM and leave the
relic density determination unaffected.

Now, we turn to justify our assumption that $\delta$ must be ${\cal O}
(1)$. If a population of KK gravitons behaves as non-relativistic
matter and decays after BBN entirely into massless (or highly
relativistic) dark radiation (DR) at a characteristic decay time $\tau_{X}$
(corresponding to redshift $z_{\rm dec}$), its contribution to the
number of extra light neutrino species $\Delta N_{\rm
  eff}$~\cite{Steigman:1977kc} at the CMB epoch  ($a = a_{\rm CMB}$) is given by \begin{equation}
\Delta N_{\rm eff} =\frac{8}{7}\left(\frac{11}{4}\right)^{4/3}\frac{\rho_{\rm DR} (a_{\rm CMB})}{\rho_{\gamma }(a_{\rm CMB})} \, ,
\label{DNeff}
\end{equation}
where the energy densities of dark radiation $\rho_{\rm DR}(a_{\rm
  CMB})$ and photons $\rho_\gamma(a_{\rm CMB})$ are defined by their distinct thermal and redshifting behaviors~\cite{Anchordoqui:2011nh}. By tracking the scaling of energy densities, $\Delta N_{\rm eff}$ can
be written in terms of the KK graviton density at the moment of
decay. Indeed, note that since KK gravitons are non-relativistic at production, for
$z_{\rm CMB} < z< z_{\rm BBN}$, their energy density redshifts like CDM
\begin{equation}
\rho_{\rm KK}(a)=\rho_{\rm KK}(\tau _{X})\left(\frac{a_{\rm
      dec}}{a}\right)^{3} \, ,
\end{equation}
where
$\rho_{\rm KK}(\tau_X)$ is the energy density of the decaying KK
gravitons right before decay.  When they decay instantly at scale factor $a_{\rm dec}$ into dark radiation, all of their energy density is converted
\begin{equation}
\rho_{\rm DR}(a_{\rm dec}) = \rho_{\rm KK}(a_{\rm dec}) \, .
\end{equation}
Once converted, this dark radiation redshifts like a relativistic fluid
\begin{equation}
\rho_{\rm DR}(a_{\rm CMB})=\rho _{\rm DR}(a_{\rm
  dec})\left(\frac{a_{\rm dec}}{a_{\rm CMB}}\right)^{4}=\rho _{\rm
  KK}(\tau _{X})\left(\frac{a_{\rm dec}}{a_{\rm CMB}}\right)^{4} \, .
\end{equation}
Photons also redshift as radiation ($\rho_\gamma \propto
a^{-4}$). Assuming no other major entropy injections happen in the SM sector during from the decay epoch to the CMB epoch we have
\begin{equation}
\rho _{\gamma }(a_{\rm CMB})=\rho _{\gamma }(\tau
_{X})\left(\frac{a_{\rm dec}}{a_{\rm CMB}}\right)^{4} \, ,
\end{equation}
where $\rho_\gamma(\tau_X)$ is the photon energy density at the time
of decay. Dividing the dark radiation density by the photon density at the CMB
epoch eliminates the scale factors entirely
\begin{equation}
\frac{\rho_{\rm DR}(a_{\rm CMB})}{\rho _{\gamma
  }(a_{\rm CMB})} =\frac{\rho
  _{\rm KK}(\tau _{X})}{\rho _{\gamma }(\tau _{X})} \, .
\label{identity}
\end{equation}
Substituting (\ref{identity}) back into (\ref{DNeff}) yields the final constant of proportionality
\begin{equation}
\Delta
N_{\rm eff} \approx
4.40\left[\frac{\rho_{\rm KK}(\tau _{X})}{\rho _{\gamma }(\tau
    _{X})}\right] \, .
\label{eNeff}
\end{equation}

To apply this calculation to a dense KK tower we have to account for the fact that the KK
gravitons are not a single particle, but a whole tower of states. As we have established in
the dark dimension scenario, intra-tower decays dominate. A heavy
graviton mode decays into lighter, still-massive graviton modes plus a
small kinetic mass-splitting fraction $\delta$. Only the kinetic
energy fraction behaves as dark radiation.  Now, if only a fraction
$\epsilon _{\rm rad}$ of the total KK energy density goes into
genuine relativistic massless modes (or high-velocity states that do
not cluster), the effective expression (\ref{eNeff}) becomes modified by that
branching ratio
\begin{equation}
\Delta N_{\rm eff}\approx 4.40\times \epsilon _{\rm rad} \times
\left[\frac{\rho _{\rm KK,\ tower}(\tau_{X})}{\rho_{\gamma }(\tau
    _{X})}\right]
\label{NeffF}
\end{equation}
Because the {\it Planck} satellite severely restricts $\Delta N_{\rm
  eff} \lesssim 0.2 \text{--} 0.3$~\cite{Planck:2018vyg}, (\ref{NeffF}) forces $\epsilon
_{\rm rad}$ to be incredibly small, which perfectly aligns with both
our assumption ($\delta \sim 1$) and the dark dimension prediction that the tower decays almost entirely into other massive dark matter states rather than dark radiation.

In closing, we note that if the KK graviton decays after BBN but
before the CMB, it decays after $e^{+}e^{-}$ annihilation is already
complete ($T \lesssim 20~{\rm keV}$). Therefore, the standard
definition of $\Delta N_{\rm eff}$ at the CMB epoch given in (\ref{DNeff}) already inherently accounts for the $e^{+}e^{-}$ entropy injection into the
photon bath. If we shift the problem such that the KK gravitons decay earlier
(overlapping with the $e^{+}e^{-}$ annihilation window at $T \sim
0.5~{\rm MeV} \to 20~{\rm keV}$), the relation (\ref{eNeff}) must be explicitly
corrected using the comoving conservation of entropy ($g_{*s} \cdot
T^3 \cdot a^3 = {\rm constant}$). In such a case, which is only relevant to $n=1$, the photon density does not redshift cleanly as $a^{-4}$ because it is being actively heated by the electron-positron plasma. The modified expression becomes
\begin{equation}
\Delta N_{\rm eff} = 4.40 \times
\left(\frac{g_{*s}(T_{\rm CMB})}{g_{*s}(T_{\rm dec})}\right)^{4/3}\frac{\rho
  _{\rm DR}(\tau _{X})}{\rho _{\gamma }(\tau _{X})}
\end{equation}
where
at $T > 0.5~{\rm MeV}$ (before annihilation),
\begin{equation}
g_{*s} = 2 \ ({\rm
  photons}) + \frac{7}{8} \times 4 \ ({\rm electrons/positrons}) =
\frac{11}{2},
\end{equation}
whereas at $T < 20~{\rm keV}$ (after annihilation), $g_{*s} =
2$. This results in a correction factor of $(4/11)^{4/3}$
which perfectly cancels out the $(11/4)^{4/3}$ factor in (\ref{DNeff}), reducing the effective multiplier down from $4.40$ to roughly $1.14$.

All in all, we conclude that the effects of KK decay are negligible and can be safely ignored from the calculation of the normalcy temperature.

\section{From 5D to 4D}\label{sec:5Dto4D}
\subsection{From 5D to the 4D Einstein frame}\label{sec:A}
In this Appendix, we present an explicit derivation of the reduction of a 5D FRW universe to its effective description in the 4D Einstein frame. Starting from the 5D action
\begin{align}\label{eq:5Daction}
   S_5=\int d^4\vec{x}dy\sqrt{-G}\left\{\frac{M_*^3}{2}{\cal R}^{(5)}-\frac{(\hat{\partial}\Phi)^2}{2}-V(\Phi)\right\}
\end{align}
with $(\hat{\partial}\Phi)^2=G^{MN}\partial_M\Phi\partial_N\Phi$, one can go to the 4D Einstein frame by writing the metric as
\begin{align}
    ds_5^2&=-d\hat{t}^2+\hat{a}(\hat{t})^2dx^2+R(\hat{t})^2dy^2\label{eq:metric1}\\
    &=\frac{R_\perp}{R(t)}\Big(-dt^2+a(t)^2dx^2\Big)+R(t)^2dy^2\label{eq:metric2}
\end{align}
where, as are interested in a 5D FRW spacetime, we write
\begin{align}\label{eq:Rtohata}
    R(\hat{t})=R_0\frac{\hat{a}(\hat{t})}{\hat{a}_0},
\end{align}
so that we obtain\footnote{We only consider the 0-modes here.}
\begin{align}
   S_E=\int d^4\vec{x}\sqrt{-g}\left\{\frac{M_p^2}{2}{\cal R}^{(4)}-\frac{(\partial\sigma)^2}{2}-\frac{(\partial\phi)^2}{2}-V_{\text{eff}}(\sigma,\phi)\right\}
\end{align}
with $(\partial\phi)^2=g^{\mu\nu}\partial_\mu\phi\partial_\nu\phi$ and the 4D Planck mass given by $M_p^2=\pi R_\perp M_*^3$. The 4D quantities are given by
\begin{align}
    R&=R_\perp\exp\left(\sqrt{\frac{2}{3}}\frac{\sigma}{M_p}\right),\\
    \Phi&=\frac{\phi}{\sqrt{\pi R_\perp}},\\
    V_{\text{eff}}&=\pi R_\perp\exp\left(-\sqrt{\frac{2}{3}}\frac{\sigma}{M_p}\right)V\left(\frac{\phi}{\sqrt{\pi  R_\perp}}\right).
\end{align}
The 5D Friedmann equations are
\begin{align}
    6H^2&=\frac{\rho_5}{M_*^3},\\
    3\hat{\dot{H}}&=-\frac{\rho_5+p_5}{M_*^3},
\end{align}
and the 5D continuity equation and the equation of motion of the inflaton $\Phi$ are given by
\begin{align}
    \hat{\dot{\rho}}_5+4H(\rho_5+p_5)&=0,\\
    \hat{\ddot{\Phi}}+4H\hat{\dot{\Phi}}+\frac{\partial V}{\partial\Phi}&=0,
\end{align}
where $H=\hat{\dot{\hat{a}}}/\hat{a}$ and the hatted dots correspond to derivatives according to $\hat{t}$, namely $\hat{\dot{}}=\partial/\partial\hat{t}$ and $\hat{\ddot{}}=\partial^2/\partial\hat{t}^2$, and where the 5D energy density and pressure are given by
\begin{align}
    \rho_5&=\frac{\hat{\dot{\Phi}}^2}{2}+V(\Phi),\\
    p_5&=\frac{\hat{\dot{\Phi}}^2}{2}-V(\Phi),
\end{align}
as we consider homogeneous solutions. The identification of the metrics \eqref{eq:metric1} and \eqref{eq:metric2} gives
\begin{align}
    \frac{\partial}{\partial\hat{t}}=\sqrt{\frac{R}{R_\perp}}\frac{\partial}{\partial t}
\end{align}
as well as
\begin{align}\label{eq:atohata}
    a=\sqrt{\frac{R}{R_\perp}}\hat{a}=\sqrt{\frac{R_0}{\hat{a}_0R_\perp}}\hat{a}^{3/2},
\end{align}
so that the 4D Hubble parameter $h$ is found to be
\begin{align}
    h=\frac{\dot{a}}{a}=\frac{3}{2}\frac{\dot{R}}{R}
\end{align}
and is linked to the 5D one through
\begin{align}\label{eq:Htoh}
    H=\sqrt{\frac{R}{R_\perp}}\left(h-\frac{1}{2}\frac{\dot{R}}{R}\right)=\sqrt{\frac{R}{R_\perp}}\frac{2h}{3}.
\end{align}
One can furhter see that the 5D equations can be re-written as the expected 4D Friedmann equations
\begin{align}
    3h^2&=\frac{\rho_4}{M_p^2},\\
    2\dot{h}&=-\frac{\rho_4+p_4}{M_p^2},
\end{align}
and to the 4D continuity equations as well as the equations of motion of the inflaton and of the radion
\begin{align}
\dot{\rho}_4+3h(\rho_4+p_4)&=0,\\
    \ddot{\phi}+3h\dot{\phi}+\frac{\partial V_{\text{eff}}}{\partial\phi}&=0,\\
    \ddot{\sigma}+3h\dot{\sigma}+\frac{\partial V_{\text{eff}}}{\partial\sigma}&=0,
\end{align}
where the 4D energy density and pressure are given by
\begin{align}
    \rho_4&=\frac{\dot{\sigma}^2}{2}+\frac{\dot{\phi}^2}{2}+V_{\text{eff}}(\sigma,\phi),\\
    p_4&=\frac{\dot{\sigma}^2}{2}+\frac{\dot{\phi}^2}{2}-V_{\text{eff}}(\sigma,\phi).
\end{align}
Finally, let's define the 5D and 4D equation of state parameters $w_5$ and $w_4$ by
\begin{align}
    p_5=w_5\rho_5\qquad\qquad\text{and}\qquad\qquad p_4=w_4\rho_4.
\end{align}
One can use the 5D and the 4D continuity and first Friedmann equations to find
\begin{align}
  \label{eq:Hwithw5}  H&=H_i\left(\frac{\hat{a}}{\hat{a}_i}\right)^{-2(1+w_5)},\\
    h&=h_i\left(\frac{a}{a_i}\right)^{-\frac{3(1+w_4)}{2}},
\end{align}
where the subscript $i$ denotes the value of the parameters at a given time, which will mostly be used at the time where the phase of constant $w$ starts. Using \eqref{eq:Rtohata}, \eqref{eq:atohata} and \eqref{eq:Htoh}, one can re-express \eqref{eq:Hwithw5} as
\begin{align}
    h=h_i\left(\frac{a}{a_i}\right)^{-\frac{4(1+w_5)+1}{3}}
\end{align}
so that we finally have 
\begin{align}
    w_4=\frac{8(1+w_5)-7}{9}.
\end{align}
Note that $w_4$ is increased compared to $w_5$, which is expected as the kinetic energy of the radion is taken into account.

\subsection{The 4D Jordan frame}\label{sec:B}
Starting from the 5D action \eqref{eq:5Daction} and decomposing the metric as
\begin{align}
    ds_5^2=G_{b,MN}dx^{M}dx^{N}+R^2dy^2,
\end{align}
the dimensionally reduced action can be written in the Jordan frame as\footnote{We only consider the 0-modes here as well.}
\begin{align}
   S_J=\int d^4\vec{x}\sqrt{-G_b}\left\{\frac{\pi R M_*^3}{2}{\cal R}^{(b)}
   -\pi R\frac{(\partial_{(b)}\Phi)^2}{2}
   -\pi RV(\Phi)\right\}.
\end{align}
Despite the absence of an explicit radion kinetic term in the Jordan frame, the radion remains a dynamical degree of freedom. Its dynamics is encoded in its non-minimal coupling to gravity through the factor multiplying the Ricci scalar. Equivalently, the effective four-dimensional Planck mass,
\begin{align}
    M_{p,J}^2=\pi R M_*^3,
\end{align}
depends on the radion and therefore evolves with time. As the size of the extra dimension increases, the effective four-dimensional Planck mass grows and the graviton becomes more weakly coupled. In contrast, the Planck mass $M_p$ appearing in the Einstein frame is fixed by construction and therefore does not directly track the evolution of the effective gravitational coupling. The radion dependence, which in the Jordan frame is encoded in the coefficient of the Ricci scalar, is made explicit after the Weyl rescaling through the radion field itself. Indeed, after performing the Weyl transformation to the Einstein frame, the radion acquires an explicit kinetic term. In the present setup, the radion also couples directly to the bulk scalar sector through both its kinetic and potential terms. 

Moreover, $G_b$ is the metric induced on the brane,
\begin{align}
    ds^2_{\rm brane}=G_{b,MN}dx^{M}dx^{N}
    =-d\hat{t}^{\,2}+\hat{a}(\hat{t})^2dx^2\,,
\end{align}
related to the Einstein-frame metric through the Weyl rescaling
\begin{align}
    G_b=\frac{R_\perp}{R}\,g\,,
\end{align}
so that
\begin{align}
    ds^2_{\rm brane}
    =\frac{R_\perp}{R}\,ds_E^2\,,
    \qquad\qquad
    ds_E^2=-dt^2+a(t)^2dx^2\,.
\end{align}
Since all matter couples to $G_b$, the metric $G_b$ is the physical metric on the brane. It is this metric, rather than the Einstein-frame metric, that determines the trajectories of freely falling observers as well as the physical times and distances measured by their clocks and rulers. Namely, free-falling clocks measure the proper time $\hat{t}$, whereas physical lengths are stretched by the scale factor $\hat{a}$.

Of course, one is free to work in whichever frame is most convenient. The Jordan and Einstein frames provide equivalent descriptions of the same physics, related by a Weyl transformation. Once the radion is stabilised, the Weyl factor becomes constant and is equal to the one in our conventions (up to small radion fluctuations), so the distinction between the two frames largely disappears.

\subsection{Dictionary between frames, dimensions and times during inflation}
In this Appendix, we further introduce the conformal time $\tau$ as
\begin{align}
    ds_5^2=\hat{a}(\tau)^2\left(-d\tau^2+d\vec{x}^2+\frac{R_0^2}{\hat{a}_0^2}dy^2\right).
\end{align}
The factors $\hat{a}_0$ and $R_0$ are the initial values of $\hat{a}$ and $R$ respectively. For the non-compact dimensions, the normalization $\hat{a}_0\equiv\hat{a}(\hat{t}_0)$ is arbitrary, but for the compact dimension, we identify $R_0$ as the physical radius of the extra dimension at initial time $\hat{t}_0$, namely $R_0\equiv R(\hat{t}_0)$, implying 
\begin{align}
    R=R_0\frac{\hat{a}}{\hat{a}_0}.
\end{align}
By defining the radius at the end of inflation $R(\hat{t}_e)\equiv R_\perp$, one finds $\hat{a}_e\equiv\hat{a}(\hat{t}_e)=\hat{a}_0R_\perp/R_0$. We further consider $\hat{a}(\hat{t})=\exp(H_I\hat{t})$, with $H_I$ is quasi-constant during inflation up to slow-roll corrections, giving a quasi-dS phase as suitable for inflation.

We further consider the metrics \eqref{eq:metric1} and \eqref{eq:metric2}:
\begin{align}
    ds_5^2&=-d\hat{t}^2+\hat{a}(\hat{t})^2dx^2+R(\hat{t})^2dy^2\\
    &=\frac{R_\perp}{R(t)}\Big(-dt^2+a(t)^2dx^2\Big)+R(t)^2dy^2,
\end{align}
and their identification lead to
\begin{align}
    a=\left(\frac{R}{R_{\perp}}\right)^{\frac{1}{2}}\hat{a}=\left(\frac{R_0}{\hat{a}_0R_{\perp}}\right)^{\frac{1}{2}}\hat{a}^{3/2},
\end{align}
giving at the beginning and at the end of inflation
\begin{align}
    a_0=\left(\frac{R_0}{R_{\perp}}\right)^{\frac{1}{2}}\hat{a}_0\qquad\text{and}\qquad a_e=\frac{R_\perp}{R_0}\hat{a}_0=\hat{a}_e
\end{align}
respectively. Therefore, as $R\sim\hat{a}$ but $a\sim\hat{a}^{3/2}$,
we understand that $\hat N$ $e$-folds in 5D correspond to $3\hat N/2$ $e$-folds in the 4D metric $ds^2_E=-dt^2+a(t)^2d\vec{x}^2$. The total numbers of $e$-folds are then found to be:
\begin{align}
    \hat{N}\equiv\ln\left(\frac{\hat{a}_e}{\hat{a}_0}\right)=\ln\left(\frac{R_\perp}{R_0}\right)
\end{align}
and 
\begin{align}\label{eq:4DefoldsAp}
    N\equiv\ln\left(\frac{a_e}{a_0}\right)=\frac{3}{2}\ln\left(\frac{R_\perp}{R_0}\right)=\frac{3}{2}\hat{N}.
\end{align}
We can now write all the relevant quantities in terms of the different times $\hat{t}$, $\tau$ and $t$. First, considering a dS phase in 5D, we have
\begin{align}
    \hat{a}(\hat{t})=e^{H_I\hat{t}},\qquad R(\hat{t})=\frac{R_0}{\hat{a}_0}e^{H_I\hat{t}}\qquad\text{and}\qquad a(\hat{t})=\left(\frac{R_0}{\hat{a}_0R_{\perp}}\right)^{\frac{1}{2}}e^{\frac{3H_I\hat{t}}{2}}
\end{align}
giving the initial and ending time of inflation to be
\begin{align}
    \hat{t}_0=\frac{1}{H_I}\ln\big(\hat{a}_0\big)\qquad
    \text{and}\qquad\hat{t}_e=\frac{1}{H_I}\ln\left(\frac{\hat{a}_0R_\perp}{R_0}\right)=\hat{t}_0+\frac{1}{H_I}\ln\left(\frac{R_\perp}{R_0}\right).
\end{align}
We can further find the relation between $\hat{t}$ and the conformal time $\tau$. As $d\hat{t}=\hat{a}d\tau$, we find at 0-th order in slow-roll
\begin{align}
    \tau=-\frac{e^{-H_I\hat{t}}}{H_I}
\end{align}
thus giving
\begin{align}
    \hat{a}(\tau)=-\frac{1}{\tau H_I},\qquad R(\tau)=-\frac{R_0}{\hat{a}_0\tau H_I}\qquad\text{and}\qquad a(\tau)=\left(\frac{R_0}{\hat{a}_0R_{\perp}}\right)^{\frac{1}{2}}\left(\frac{-1}{\tau H_I}\right)^{\frac{3}{2}}
\end{align}
as well as
\begin{align}
    \tau_0=-\frac{1}{\hat{a}_0H_I}\qquad\qquad\text{and}\qquad\qquad\tau_e=-\frac{R_0}{\hat{a}_0H_IR_\perp}=\tau_0\frac{R_0}{R_\perp}.
\end{align}

We then do the same analysis for the time $t$. Using $\hat{a}d\tau=(R_{\perp}/R)^{1/2}dt$, we find at leading order in slow-roll
\begin{align}
    t=\frac{2}{H_I}\left(\frac{R_0}{\hat{a}_0R_{\perp}}\right)^{\frac{1}{2}}\left(\frac{-1}{\tau H_I}\right)^{\frac{1}{2}}=\frac{2}{H_I}\left(\frac{R_0}{\hat{a}_0R_{\perp}}\right)^{\frac{1}{2}}e^{\frac{H_I\hat{t}}{2}}
\end{align}
so that
\begin{align}
    \hat{a}(t)=\frac{R_{\perp}}{R_0}\left(\frac{H_I t}{2}\right)^2\hat{a}_0,\qquad R(t)=R_{\perp}\left(\frac{H_I t}{2}\right)^2,\qquad\text{and}\qquad a(t)=\frac{R_{\perp}}{R_0}\left(\frac{H_I t}{2}\right)^3\hat{a}_0
\end{align}
as well as
\begin{align}
    t_0=\frac{2}{H_I}\left(\frac{R_0}{R_{\perp}}\right)^{\frac{1}{2}}\qquad\text{and}\qquad t_e=\frac{2}{H_I}=t_0\left(\frac{R_\perp}{R_0}\right)^{\frac{1}{2}}.
\end{align}
We also introduce the 4D Hubble parameter $h$ defined by
\begin{align}
    h\equiv\frac{\dot{a}}{a}=\frac{3}{t}
\end{align}
where the dot stands for a derivative with respect to $t$. We further have
\begin{align}
    h_0=\frac{3H_I}{2}\left(\frac{R_{\perp}}{R_0}\right)^{\frac{1}{2}}\qquad\text{and}\qquad h_e=\frac{3H_I}{2}=h_0\left(\frac{R_0}{R_\perp}\right)^{\frac{1}{2}}
\end{align}
while the comoving horizons are linked through
\begin{align}\label{eq:comovinghorizons}
    ah=\frac{3}{2}\hat{a}H_I.
\end{align}

\begin{table}[!h]
    \centering
    \begin{tabular}{c|C{2.5cm}|C{2.5cm}|C{2.5cm}|}
    \cline{2-4}
         & 5D & 4D Jordan & 4D Einstein \\
         \hline
        \multicolumn{1}{|c|}{scale factor}   & $\hat{a}$ & $\hat{a}$ &$a$\\
        \multicolumn{1}{|c|}{Hubble parameter}   & $H_I$ & $H_I$ &$h$\\
        \multicolumn{1}{|c|}{cosmic time}   & $\hat{t}$ & $\hat{t}$ &$t$\\
        \multicolumn{1}{|c|}{number of e-folds}   & $\hat{N}$ & $\hat{N}$ &$N$\\
        \multicolumn{1}{|c|}{Planck mass}   & $M_*$ & $\sqrt{\pi R M_*^3}$ &$M_{p}$\\
        \hline
    \end{tabular}
    \caption{Dictionary between the frames}
    \label{tab:dict}
\end{table}

 \section{The transition scale}\label{sec:lambda_t}
 In this Appendix, we derive the value of the transition wavelength today from the analysis \cite{Petretti:2024mjy}. As fixing the normalisation of the scale factor has to be done carefully in the case of 5D inflation, we will not assume any normalisation here. Indeed, the authors of \cite{Petretti:2024mjy} use the usual convention $a_t=1$ while using the results of \cite{Antoniadis:2023sya} the convention used is $\hat{a}_0=1$. However, we will see that the choice of normalisation does not affect the physical results of the analysis.

 Unpacking the normalization of $\hat{a}$ actually consists in replacing $R_0$ by $R_0/\hat{a}_0$. Following \cite{Petretti:2024mjy}, we therefore introduce the comoving momentum $k_{\text{DD}}$ as
 \begin{align}
     \frac{R_0}{\hat{a}_0}\equiv\frac{1}{\pi k_{\text{DD}}}.
 \end{align}
 Our condition of scale invariance
 \begin{align}\label{eq:ksi}
     \frac{\pi k R_0}{\hat{a}_0}>1 
 \end{align}
 is then $k>k_{\text{DD}}$. This scale can be linked to a chosen pivot scale $k_*$ as 
\begin{align}
    k_{\text{DD}}=k_*e^{N_{\text{DD}}-N_*}
\end{align}
where $N_{\text{DD}}$ and $N_*$ are respectively the numbers of e-folds since the beginning of inflation at which $k_{\text{DD}}$ and $k_*$ exit the horizon. The pivot scale $k_*$ chosen in the analysis \cite{Petretti:2024mjy} is the usual $k_*/a_t=0.05\,\text{Mpc}^{-1}$, where we re-introduce the scale factor today $a_t$. Using \eqref{eq:ksi}, the physical transition wavelength today is then given by
\begin{align}
    \lambda_t=2\pi^2a_t\frac{R_0}{\hat{a}_0}=\frac{2\pi a_t}{k_{\text{DD}}}=\frac{2\pi a_t}{k_*e^{N_{\text{DD}}-N_*}}.
\end{align}
The authors of \cite{Petretti:2024mjy} then find $N_\text{DD}-N_*<-5.38$ at $95\%$ confidence, giving
\begin{align}
    \lambda_t>27.3\,\text{Gpc}.
\end{align}

\begin{figure}[!htbp]

\centering

\begin{subfigure}{0.45\textwidth}
\centering
\begin{tikzpicture}
\draw[->] (0,0) -- (6,0);
\draw[->] (0,-2) -- (0,2);

\def\N{2.5}

\draw[dashed] (1.2,0) -- (1.2,1.2);
\draw[dashed] (3.6,0) -- (3.6,1.2);
\draw[dashed] (3,0) -- (3,-1.2);
\draw[dashed] (4.2,0) -- (4.2,-1.2);

\draw[purple, thick, domain=0:5.9, samples=200]
plot (\x,{sin(deg(2*pi*\N*\x/6))});

\node[circle,draw=red,fill=red!30,inner sep=2pt] at (0.6,0) {};
\node[text=red,below] at (0.6,0) {hot};

\node[circle,draw=blue,fill=blue!30,inner sep=2pt] at (1.8,0) {};
\node[text=blue,above] at (1.8,0.1) {cold};

\node[circle,draw=red,fill=red!30,inner sep=2pt] at (3,0) {};
\node[text=red,below] at (3,0) {hot};

\node[circle,draw=blue,fill=blue!30,inner sep=2pt] at (4.2,0) {};
\node[text=blue,above] at (4.2,0.1) {cold};

\node[circle,draw=red,fill=red!30,inner sep=2pt] at (5.4,0) {};
\node[text=red,below] at (5.4,0) {hot};

\draw[<->, thick] (1.2,1.2) -- (3.6,1.2)
node[midway, above] {$\lambda$};

\draw[<->, thick] (3,-1.2) -- (4.2,-1.2)
node[midway, below] {$\frac{\lambda}{2}=\theta D_{\text{LSS}}$};
\end{tikzpicture}
\caption{}
\end{subfigure}
\hfill
\begin{subfigure}{0.45\textwidth}
\centering
\begin{tikzpicture}

\def\R{3}      
\def\A{0.4}   

\draw[black] (0,0) circle (\R);

\draw[very thick,purple,samples=200,domain=15:75]
plot ({(\R+\A*sin(6*\x-270))*cos(\x)},
     {(\R+\A*sin(6*\x-270))*sin(\x)});

\draw[<->,thick]
(30:\R+0.55) arc (30:60:\R+0.55);

\node at (42:\R+1.1) {$\theta D_{\text{LSS}}$};

\draw[-]
(30:\R-2.5) arc (30:60:\R-2.5);

\node at (45:\R-2.25) {$\theta$};

\draw[dashed] (0,0) -- (30:3);
\draw[dashed] (0,0) -- (60:3);
\draw[<->, thick] (-3,0) -- (0,0)
node[midway, below] {$D_{\text{LSS}}$};

\node[circle,draw=red,fill=red!30,inner sep=2pt] at (60:\R) {};

\node[circle,draw=blue,fill=blue!30,inner sep=2pt] at (30:\R) {};
\end{tikzpicture}
\caption{}
\end{subfigure}

\caption{Relation between the wavelength of a primordial fluctuation and the
angular scale observed in the CMB. Left: a fluctuation of wavelength $\lambda$
produces alternating hot and cold spots; the observed hot--cold separation
corresponds to half of the wavelength. Right: projection of this pattern on the
Last Scattering Surface at distance $D_{\rm LSS}$, showing that an angular scale
$\theta$ on the sky probes a physical scale $\theta D_{\rm LSS}$.}
\label{fig:hotcold}
\end{figure}
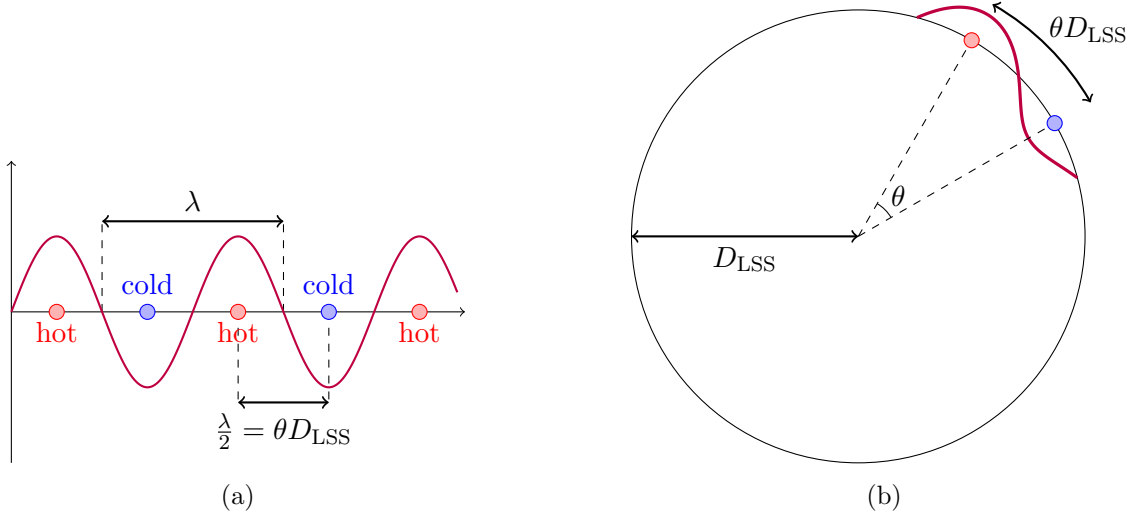

It is often convenient to express these scales in terms of the CMB multipole 
$\ell$, related approximately to the physical wavenumber $k$ of primordial 
fluctuations through $\ell \simeq k D_{\rm LSS}$, where $D_{\rm LSS}$ denotes the proper distance to the Last Scattering Surface. It is given today by $D_{\rm LSS} \simeq 14\,{\rm Gpc}$, so that the transition multipole is
\begin{align}
    \ell\simeq\frac{2\pi}{\lambda_t}D_{\rm LSS}\lesssim3.23.
\end{align}
The discrepancy with the $\ell = 7$ estimate
of~\cite{Anchordoqui:2024amx} arises from the $a_0 R_0 \approx
R_{\perp}$ assumption. As discussed in Sec.~\ref{setup}, this assumption leads to
a discontinuity when matching the scale factor's extrapolation (from
the present back into the past) with the end of 5D inflation. Now,
$\ell \lesssim 3.23$ further corresponds to the angular scale $\theta \sim \pi/\ell$, which can be directly connected to the transition wavelength through
\begin{equation}
\lambda_t \sim 2 \theta D_{\rm LSS}\qquad\qquad\text{giving}\qquad\qquad\theta\gtrsim 56^\circ.
\end{equation}
The factor of two can also be understood heuristically. A primordial 
sinusoidal fluctuation produces alternating hot and cold regions in the CMB, 
so the angular size of a hot--cold pair corresponds roughly to half of the 
underlying wavelength, as illustrated in Fig.~\ref{fig:hotcold}.\footnote{The heuristic argument implicitly treats primordial perturbations as sinusoidal plane waves projected onto a locally flat patch of the LSS. The exact relation between the multipole $\ell$ and the physical wavenumber $k$ instead arises from spherical Bessel functions $j_\ell(kD_{\rm LSS})$ in the CMB transfer functions. For large multipoles these functions peak sharply near $kD_{\rm LSS}\sim\ell$, reproducing the flat-sky intuition, while at low multipoles the projection becomes broad and the correspondence between angular scale and physical wavelength becomes increasingly imprecise.}

\bibliography{biblio}

\end{document}